Behavioral and EEG evidence for inter-individual variability in late encoding stages of word production

Pamela Fuhrmeister*[1]

Sylvain Madec[1]

Antje Lorenz[2]

Shereen Elbuy[1]

Audrey Bürki[1]

[1] Department of Linguistics

University of Potsdam

Karl-Liebknecht-Straße 24-25

14476 Potsdam, Germany

[2]Department of Psychology, Neurocognitive Psychology

Humboldt-Universität zu Berlin

Unter den Linden 6

10099 Berlin, Germany

*Corresponding author, email: fuhrmeister@uni-potsdam.de

Abstract

Individuals differ in the time it takes to produce words when naming a picture. However, it is unknown whether this inter-individual variability emerges in earlier stages of word production (e.g., lexical selection) or later stages (e.g., articulation). The current study measured participants' ($N = 45$) naming latencies and continuous EEG in a picture-word-interference task, as well as naming latencies in a delayed naming task. The inter-individual variability in naming latencies in immediate naming was not larger than the variability in the delayed task. Thus, a large part of the variability in immediate naming seems to originate in relatively late stages of word production. This interpretation was complemented by the EEG data: Differences between relatively fast vs. slow speakers were seen in response-aligned analyses in a time window close to the vocal response. Finally, we show that inter-individual variability can influence EEG results at the group level.

Behavioral and EEG evidence for inter-individual variability in late encoding stages of word production

The aim of psycholinguistics is to describe the cognitive system that allows individuals to produce and comprehend language. Work in the field (as in cognitive psychology in general) rests on an important assumption, which is reflected in the methodological approach. This assumption is that a common (universal) architecture is reflected in the group behavior. Inter-individual differences in behavior are acknowledged, but the assumption is that the common architecture will be visible when filtering out the noise, i.e., inter-individual differences. In the present study, we take a different stance. We join a marginal but perhaps increasing number of studies that examine how inter-individual differences contribute to the understanding of word production. In addition, we ask whether inter-individual differences may in some cases prevent us from accessing this common architecture.

We bring together two literatures, both in the field of language production, that make opposite assumptions about inter-individual differences in the time course of language production processes. The first assumes that speakers differ in the time they need to perform specific encoding processes (e.g., Jongman et al., 2015; Laganaro et al., 2012; Shao et al., 2012). The second assumes (albeit implicitly) that encoding processes are synchronous enough across speakers, such that the underlying processes can be targeted with time course measures, e.g., event-related potentials (ERPs, e.g., Bürki, 2017b; Krott et al., 2019; Rabovsky et al., 2020). The aim of this study is to determine where in the time course of word production these inter-individual differences in naming latencies (time to prepare the vocal response to name a picture of an object) emerge. The functional origin of inter-individual differences in word production processes has crucial theoretical and methodological implications. These implications, as well as relevant literature, are discussed below.

**Theoretical implications of individual differences in word production**

Although speaking often seems effortless (at least for adults without language impairment), it is not a simple task. According to prominent models of word production, the process of producing a single word is made up of several stages: conceptualization, lexical selection, phonological encoding, phonetic encoding and execution of articulatory gestures (e.g., Indefrey & Levelt, 2004; Indefrey, 2011; Levelt, 1999). Producing words is not entirely an automatic process, and it is at least partially subject to cognitive resources (e.g., attentional control, Piai et al., 2013). Authors typically assume that the encoding processes involved in word production differ in their degree of automaticity, and as a consequence, in the extent to which they rely on available cognitive resources. The lack of automaticity of (a subset of) encoding processes and their reliance on cognitive resources has often been linked to the variability observed in the time needed to encode a vocal response in the naming task. This variability is substantial. In our own work, we have observed mean naming latencies that have ranged between 613 and 1070 ms in a simple picture naming experiment (Valente et al., 2014) and between 615 to 924 ms in a picture-word interference task (Bürki, 2017a). The large standard deviations reported in most studies corroborate our observations that variability in naming times across participants is substantial.

If certain stages of word production proceed less automatically and instead are susceptible to influences from cognitive resources, we would expect to observe inter-individual differences in these stages, for example, in the time it takes an individual to select a word during the lexical access stage (see for instance Laganaro et al., 2012). On the other hand, if some stages of word production proceed automatically, we would not expect much individual variability to emerge in these stages. Understanding this variability (and being able to locate it in the speech production system) will inform us on crucial aspects of this system:

its links with cognitive resources, autonomy of the language system, and the degree of automaticity in word production processes.

There is currently no consensus as to which processes or stages of word production are more or less automatic, despite the number of studies that have addressed this question (see (Garrod & Pickering, 2007; Hartsuiker & Moors, 2017; Jongman et al., 2015; Meyer et al., 2007; Roelofs, 2008). Some studies have examined individual differences in picture-naming times by testing whether cognitive skills predict response times at different stages of word production. For example, Jongman et al. (2015) examined the link between sustained attention (i.e., defined as "the ability to maintain alertness over time") and picture-naming times. In one experiment, participants performed only a picture-naming task, and in a second experiment, participants performed a dual task. The authors observed a correlation between the tail of the naming time distribution and a measure of sustained attention in both experiments (with a higher correlation in the dual task context), but no correlation between sustained attention and gaze duration. Considering previous research showing that gaze duration reflects all planning stages up to the phonological encoding process (Meyer et al., 1998), they conclude that sustained attention is mostly involved after the phonological encoding process (presumably phonetic encoding and execution processes), and therefore that late stages of word production do not necessarily proceed automatically. Note that this interpretation is based on the absence of a correlation between gaze duration and the sustained attention measure (and see Jongman et al., 2015 for contrasting results for production of conjoined noun phrases). Shao et al. (2012) found that various measures of executive control (updating and inhibiting) predicted picture naming reaction times, but they found stronger associations between executive control and naming times when participants named actions compared to objects. They argue that updating and inhibiting are involved in a pre-lexical, conceptual stage of word production because their stimuli (object and action pictures) were well matched in terms of lexical attributes and because actions are conceptually more complex than objects.

Other studies have used the picture-word interference paradigm to investigate when in the different stages of word production individuals vary. This paradigm is useful for testing this question because its effects are often associated with specific encoding processes. In the picture-word interference task, participants see a picture presented on a screen with either a written distractor item superimposed on the picture or a spoken word played via headphones and are asked to name the picture and ignore the distractor word. The typical effects that are associated with this paradigm include a general interference effect, a phonological facilitation effect, and a semantic interference effect. The general interference effect describes the effect that is seen when distractor words that are superimposed on the picture lead to slower response times than strings of non-meaningful symbols (e.g., a line of Xs; Bürki, 2017b; La Heij & Vermeij, 1987a; Lupker, 1982). In other words, seeing a written word interferes more with naming a picture than seeing for instance a line of Xs.

When the target (the picture to be named) and distractor overlap in their spelling or phonology, the target is produced faster than when the two words are unrelated (i.e., phonological facilitation effect, Jescheniak et al., 2003; Posnansky & Rayner, 1977; Rayner & Posnansky, 1978; Damian & Martin, 1999; Starreveld & La Heij, 1996; Bi, Xu, & Caramazza, 2009; De Zubicaray, McMahon, Eastburn, & Wilson, 2002). This effect has been associated with the phonological encoding stage of word production, which, according to the dominant view in the field, is initiated after lexical selection but before articulation (e.g., Indefrey & Levelt, 2004; Meyer, 1996; Meyer & Schriefers, 1991; see also Lupker, 1982; Schriefers et al., 1990; Starreveld & La Heij, 1995; Starreveld & La Heij, 1996; Starreveld, 2000; Zhang & Weekes, 2009).

It has further been shown that distractors of the same semantic category as the target word (target = kiwi, distractor = pineapple) produce interference, i.e., are associated with slower response latencies, compared to semantically unrelated words, which is referred to as the semantic interference effect. This effect has predominantly been associated with lexical access (e.g., Levelt, 1999), but this locus is debated and other authors associate it with pre-execution stages (e.g., Mahon et al., 2007). According to the response-exclusion hypothesis, the distractor word is temporarily stored in a pre-articulatory response buffer. In order to produce the target word, the distractor word needs to be removed from this buffer, and the semantic interference effect arises because related distractors are more difficult to discard than unrelated ones (Mahon et al., 2007). A few authors provide evidence that the semantic interference effect could have two loci or show evidence that the effect is associated with both lexical selection and a post-lexical articulatory stage (Krott et al., 2019).

Many studies have used the picture-word interference paradigm to link individual variability in cognitive skills to specific stages of word production and have found evidence of individual variability in lexical access. For example, Ferreira & Pashler (2002) used a dual-task paradigm, in which participants completed a picture-word interference task and a tone discrimination task. In this paradigm, two tasks are performed sequentially with varied stimulus onset asynchronies of the stimuli presented for each task. It is assumed that if performance on both tasks share some processing resources, the time it takes to complete the second task will increase if the duration between the presentation of the two stimuli is shorter. This is because central attention mechanisms are first allocated to the first task, and then to the second task. Ferreira and Pashler (2002) found that response times in the tone discrimination task did not differ between phonologically related and unrelated trials in the picture-word interference task but were influenced by the semantic contrast. Therefore, they concluded that lexical selection, but not phonological encoding, involves attentional resources. However, this interpretation relies on the assumption that the semantic interference effect arises during lexical selection. Other studies have used a similar paradigm to determine the locus of the semantic interference effect and have obtained mixed results. For example, in studies by Dell et al. (2007) and Ayora et al. (2011) effects of semantic interference were found at long, but not short stimulus onset asynchronies, which they take as evidence that the semantic interference effect originates in the conceptual stage (i.e., prior to lexical selection). Other studies that used this paradigm found semantic interference effects regardless of the stimulus onset asynchrony, which they interpret as support for a lexical selection or later locus of the semantic interference effect (Piai et al., 2014; Schnur & Martin, 2012). Therefore, it is still unclear whether lexical selection relies on attentional resources.

In addition to the behavioral studies discussed above, a couple of studies have used EEG to address the question of individual variability in the time course of word production. For example, Shao et al. (2014) compared naming latencies between pictures of low and high name agreement (a measure of how often people assign the same name to a picture). A measure of selective inhibition, computed for each individual, correlated negatively with the difference in amplitude of the EEG signal between trials with low and high name agreement in the 170-330ms time window after picture onset. They therefore associate inhibition with the lexical selection phase and suggest that inhibition allows participants to reduce competition between lexical items, as pictures with lower name agreement have more possible names to choose from. Laganaro et al. (2012) compared ERPs, both aligned on the picture presentation and on the vocal response, of slow and fast speakers. They found differences across speed groups only between 200 and 300ms after picture onset, a time window they relate to lexical selection (but see below).

Taken together, these findings confirm that word production is not an entirely automatic process and that participants vary in the time it takes them to prepare a linguistic response for production. However, the locus of this variability is not yet well understood, and we still do not have a complete picture of where in the process of word production participants are most variable. In most studies, the link between variability observed in the behavior of participants and a specific encoding process requires additional assumptions (e.g., semantic interference reflects lexical access, gaze durations reflect phonological encoding, name agreement effects only arise during lexical access). More research is clearly needed to clarify this issue. As discussed in the next section, the locus of this inter-individual variability may have important methodological consequences for studying the time course of word production.

**Methodological implications of individual variability in the time course of word production**

In the last ten years, an increasing number of studies have used measurements with high temporal resolution, such as MEG or EEG, to examine word production processes (e.g., Ganushchak et al., 2011; Munding et al., 2015; Strijkers et al., 2010). Several contemporary issues require fine grained information about the time course of events, i.e., how the activation flows in the system or the degree of seriality of encoding processes (e.g., sequentially vs. parallelism in underlying encoding processes, Munding et al., 2015; Riès, 2016; Miozzo et al., 2014; Costa et al., 2009). Measures with high temporal precision, such as EEG or MEG, can be used for these purposes. However, temporal precision in EEG and MEG can only be achieved if neural activity is synchronous across participants. The traditional approach to analyze results from ERP experiments involves averaging dependent variables across participants. This approach allows filtering instrumental noise and inter-individual differences. Observed differences in response times, together with claims that encoding processes involve general domain cognitive resources whose availability varies among individuals (see above), are difficult to reconcile with this synchronicity assumption.

As reviewed above, several studies argue that participants vary in the time they need to access lexical representations. This claim has crucial implications for EEG studies. The cognitive processes underlying language production are not signaled by well-known components, as can be the case in other fields (e.g., N400 and P600 in language comprehension, N170 in face processing). As a result, the time course of experimental effects is often taken as an indicator of underlying encoding processes. For instance, the estimates of the time course (onset, duration) of encoding processes (i.e., conceptualization, lexical access, phonological encoding) provided by Indefrey (2011) or Indefrey and Levelt (2004) in their meta-analyses are often used to map experimental effects with underlying cognitive processes. According to these authors, in an experimental setup where participants need to produce the name of a picture, assuming several repetitions of the same word and a reaction time of 600 ms, it is estimated that visual recognition and conceptualization occur up until 175-200 ms after visual onset. Lemma selection then takes place from 200 ms until about 275 ms and is followed by encoding of the phonological form between 275 and 450 ms after picture onset. Finally, phonetic encoding occurs around 450-600 ms, until the initiation of motor execution (Indefrey & Levelt, 2004). Many studies use this time course to map a given effect to a given encoding process (e.g., the effect arises at around 250 ms, it is therefore associated with lexical access). If participants indeed vary in the time they need to complete lexical access (or phonological encoding), this general time course cannot be assumed to generalize across studies that use different groups of participants. Some researchers rescale this time frame proposed by Indefrey and Levelt (2004) to fit longer or shorter response times (see for example Shitova et al., 2017). To accurately rescale these time frames, however, we need to determine whether faster speakers are faster in all stages of word production (e.g., Schuhmann et al., 2009) or whether faster speakers are faster in only some stages of word production (e.g., Laganaro et al., 2009). Establishing exactly which stages of word production are affected by inter-individual variability will help determine which stages' time frames should be rescaled and by how much.

   More generally, if differences in response times are due to differences in the duration of lexical access across participants, stimulus-aligned ERPs can hardly be used to study encoding processes that occur after lexical access, as the ERPs monitored during these processes will no longer be aligned across participants. Given these considerations, there is an urgent need to establish to what extent participants vary in the timing of (which) encoding processes. As stated above, the aim of this study is precisely to contribute to this endeavor.

**Current study**

In the current study, we seek to determine where in the time course of word production inter-individual differences emerge. To test this behaviorally, participants performed a picture-word interference task and a delayed naming task. EEG activity was monitored only during the picture-word interference task.

We have argued that individual variability in reaction time that has been observed in many studies of word production does not allow us to generalize the time course of pre-articulatory planning stages proposed by (Indefrey & Levelt, 2004) to all participants. This may be especially true if inter-individual variability arises in early stages of word production such as lexical selection. In light of this discussion, a more precise test of the ERP markers of underlying processes and their time course could be achieved by manipulating experimental conditions of a picture naming task that have been shown to induce the semantic interference and phonological facilitation effects (see Bürki et al., 2020 for review). As discussed above, it has been argued that the phonological facilitation effect occurs during the phonological encoding stage (Indefrey & Levelt, 2004) and the semantic interference effect during lexical selection (Levelt et al., 1999) and/or pre-articulatory processes (Mahon et al., 2007; see also Bürki et al., 2020 for review). Therefore, these task manipulations could serve as markers of the specific encoding processes of interest, rather than relying on the proposed time windows of these encoding processes, as these may not be generalizable across participants. In the present study, we use the phonological facilitation effect as a marker of phonological encoding, and the semantic interference effect as a marker of lexical access and/or later pre-articulatory processes, depending on the time course of the effect at the group level. The ERP effect will be interpreted mainly as reflecting lexical access if it occurs in the stimulus-aligned ERPs, shortly after picture presentation, and before the phonological facilitation effect. On the other hand, the effect will be interpreted as reflecting later pre-articulatory processes if it is observed in the response-aligned ERPs, close to the onset of articulation, and after the phonological facilitation effect.

We analyze experimental effects in EEG at the group level and perform several analyses (planned and exploratory)[1] to determine how experimental effects vary with participants' response speed. In addition, we compare the inter-individual variability in naming times in immediate and delayed naming tasks. Naming times in immediate naming tasks are thought to reflect all processes up to execution of gestures, whereas naming times in delayed task are thought to reflect only stages after phonetic encoding (e.g. Laganaro & Alario, 2006). By comparing the variability in the two tasks, we obtain information on the amount of variability that arises before /after articulatory execution. We reasoned that if the between-participant variability that we observe in the immediate naming times originates in earlier processes (i.e., mapping between the picture and the concept, lexical access, or phonological encoding), between-participant variability should be much greater in the immediate task than in the delayed task. Moreover, the by-participant mean naming times in the two tasks should not be correlated. By contrast, if a large part of the variability in response times in the immediate naming task reflects late execution processes, we would expect the variability in the two tasks to be similar, and we would expect to see a relationship between the by-participant mean naming times in the two tasks.

---

[1] Our initial plan was to perform single participant analyses of experimental effects from the EEG data and correlate onset/offset of effects with response times. This could not be done because too few participants showed effects in the EEG data after correction for multiple comparisons. It is possible that the variability across items was too high. For instance, the picture with the fastest mean response time was produced 270 ms faster than the picture with the slowest mean response time. We also wanted to explore the relationships between this timing and cognitive resources. Information on general domain cognitive tasks and correlations with naming times are presented in supplementary materials at https://osf.io/svjh5/. In short, we find that naming times are modulated by a measure of sustained attention. However, given the number of different measures related to cognitive resources that we used, this relationship could be spurious. We are currently working on a replication of these findings.

If participants vary in the duration of one or more stages of word production, this might compromise the use of EEG to target specific encoding processes. For example, if variability is found in lexical selection, EEG may not be an appropriate tool to analyze this data, as participants' neural activity will not be synchronized after the presentation of a stimulus. On the other hand, if variability is found in later encoding stages, EEG may be able to be used to examine effects at earlier encoding stages.

## Method

### *Participants*

Forty-five right-handed, native speakers of German (ages 18-30, mean = 23.2, SD = 3.5) took part in the experiment. None reported hearing, psychiatric, neurological or linguistic disorders, and their participation was rewarded either by course credit or money. Participants were given details about the experimental procedure and provided their informed consent prior to participation. The study received ethical approval by the Ethical committee of the University of Potsdam (Germany).

### *General procedure*

Each participant was tested in three different sessions. In the first session, participants completed three tasks: picture-word interference, delayed naming, and word naming (reading aloud). EEG activity was monitored only while participants performed the picture-word interference task. The reading aloud task is not reported here but is described in a companion article. In the second and third sessions, participants performed a series of standardized cognitive tests that are also not reported here (see supplementary materials for more information at https://osf.io/svjh5/).

*Picture-word interference task*

*Material*

We selected 90 German nouns (thereafter "target words") and their corresponding object pictures from the Multipic database (see Appendix 1; Duñabeitia et al., 2017). Lemma frequencies for these words, according to the *dlex* database (Heister et al., 2011), ranged from 9 to 21184 (mean = 1991.3, SD = 3678.4). In addition, we selected 180 nouns to be used as distractors. A total of 450 stimuli were created by combining the 90 pictures with the distractor words or a line of Xs, so as to form the following five conditions: (1) Baseline: the 90 pictures were displayed synchronously with a series of 6 Xs, (2) semantically related: each picture was combined with a distractor noun from the same semantic category (e.g., two animals, two tools), (3) semantically unrelated: this condition was created by assigning the 90 distractors from the semantically related condition to a different picture such that picture and distractor had no semantic or phonological relationship, (4) phonologically related: each picture was displayed with a noun overlapping in the first phoneme(s) with the target word (between 1 and 4 phonemes, mean = 2.62, S.D: 0.67), (5) phonologically unrelated: this condition was created by assigning the 90 distractors from the phonologically related condition to a different picture such that picture and distractor had no semantic or phonological relationship.

The distractor was superimposed on the picture and was written in white with a black outline in Arial font. Stimulus-onset asynchrony was 0 ms. The first letter was uppercase because German nouns are always capitalized. Pictures and distractors were presented against a gray background. Figure 1 displays the target word banana (*Banane* in German) with its distractors. Eight additional pictures were selected as training or filler items and associated with 16 new distractor words following the same criteria as for the test items.

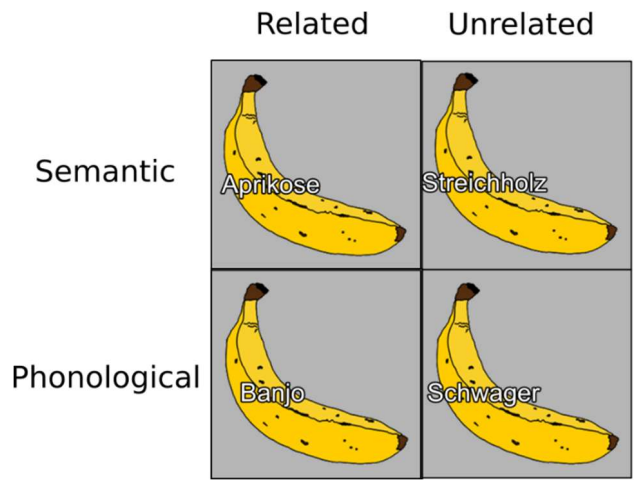

Figure 1. *Example of stimuli in the picture-word interference task. The target word "Banane" (banana) with a semantically related (Aprikose/apricot), semantically unrelated (Streichholz/match), phonologically related (Banjo/banjo), and phonologically unrelated (Schwager/brother-in-law) distractor. Note that two unrelated conditions were used so as to have the same distractor lists when contrasting a related and unrelated condition.*

*Task description*

The picture-word interference task started with a familiarization phase. During this phase, pictures (without superimposed distractors) were presented one by one on the screen, together with their name. Participants were instructed to be attentive to the pictures, and to silently read the corresponding word. The task was self-paced; the items were presented in random order.

Next, participants completed the main picture-word interference task. Participants were told to name the pictures displayed on the computer screen as fast and accurately as possible. The task consisted of a short training phase followed by five blocks of test trials. Each target word appeared only once within each experimental block; order of presentation was pseudorandomized in each block, and the number of trials in each condition was equated in each block. Each participant received a different order. Each block started with four filler trials. A trial started with a fixation cross, whose duration ranged between 2200 and 2300ms. The picture-word pair was then displayed at the center of the screen for maximally 2300ms or until a response was given (visual angle of the square: 5.5°, vertical visual angle of the text: 0.6°). Vocal responses were recorded starting from the picture-word onset until 3000ms after picture onset. The inter-trial interval was a random duration between 1000 and 1200 ms. Continuous EEG was recorded during this task.

*Delayed naming task without distractors*

Stimuli for the delayed naming task consisted of the same 90 test pictures and 8 additional filler pictures used for the picture-word interference task. A trial had the following structure: A fixation cross first appeared at the center of the screen. After 700ms, it was replaced by the picture (without a distractor or line of Xs). The picture stayed on the screen for 1000ms[2], and participants were instructed to wait until a response cue (a blue circle) appeared before naming the picture aloud as quickly as possible. Next, a blank screen replaced the picture for a random duration ranging from 1200ms to 2000ms. Then a blue circle (the response cue) appeared in the middle of the screen and stayed there for 1500ms. The next trial started after a 300ms inter-trial interval. The experiment started with eight training items followed by four filler items. The 90 test items were then presented one by one, in random order.

**EEG acquisition and preprocessing**

Continuous EEG was recorded with a BrainAmp MR amplifier. Sixty one Ag/Ag-Cl electrodes were positioned according to the extended 10-system: Fp1, Fp2, AF7, AF3, AF4,

---

[2] This duration is shorter than it was for the picture-word interference task. If these durations had been equated between the two tasks, the trials for the delayed naming task would have been much longer due to the interval between picture presentation and the response cue.

AF8, F7, F5, F3, F1, Fz, F2, F4, F6, F8, FT7, FC5, FC3, FC1, FCz, FC2, FC4, FC6, FT8, T7, C5, C3, C1, Cz, C2, C4, C6, T8,TP7, CP5, CP3, CP1, CPz, CP2, CP4, CP6, TP8, P7, P5, P3, P1, Pz, P2, P4, P6, P8, PO9, PO7, PO3, POz, PO4, PO8, PO10, O1, Oz and O2. FCz was used as the reference electrode during the recording. The sampling rate was set to 1000 Hz.

The EEG signal was preprocessed in MATLAB, using the EEGLAB toolbox (Delorme & Makeig, 2004). Continuous signals were re-sampled at 500 Hz and recomputed against the average of all electrodes. Data were then filtered with a 0.1 Hz high pass filter (Kaiser windowed sinc FIR filter, order = 8008, beta = 4.9898, transition bandwidth = 0.2 Hz) and a 40 Hz low pass filter (Kaiser windowed sinc FIR filter, order = 162, beta =4.9898, transition bandwidth = 10 Hz). Continuous EEG signals were segmented into epochs of 3.4s, from -1s before visual stimuli onsets to 3.4s after onsets. These epochs were manually inspected, and noisy channels were spherically interpolated (mean of channel interpolated by participant = 0.7, range = [0-5]), while noisy epochs excluded (mean of epochs rejected by participant = 10.5, range = [0-47]). Artefacts corresponding to blinks were corrected using independent components analysis (ICA, Chaumon et al., 2015). In order to improve the ICA decomposition, a second data set was created, differing solely from the first one by using a 1 Hz high pass filter (Kaiser windowed sinc FIR filter, order = 802, beta = 4.9898, transition bandwidth = 2 Hz). The obtained demixing matrix was then applied to the data set filtered at 0.1 Hz. Individual components corresponding to blinks were then excluded in a semi-automatic way, relying on the SASICA plugin and on manual inspection. After removal of blink components, a second step of data cleaning was performed in a semi-automatic way. For each epoch, we detected channels presenting an amplitude superior to +100 μV or inferior to -100 μV. If a channel was detected on more than 45 epochs, the channel was spherically interpolated on every epoch. If an epoch had more than three channels detected, it was disregarded. If an epoch had less than three channels detected, we spherically interpolated these channels for the epoch duration. Then we screened epochs for channels with abnormal

INDIVIDUAL VARIABILITY IN LATE STAGES OF WORD PRODUCTIONtrends (*rejtrend* function; slope > 50 µV with R2 > 0.3). If an abnormal trend was detected on more than 45 epochs, the channel was spherically interpolated on every epoch. If an epoch had more than three channels detected, it was disregarded. If an epoch had less than three channels detected, we spherically interpolated these channels for the epoch duration. Finally, we performed a last manual check of the remaining epochs and excluded the remaining noisy channels or electrodes. Finally, we excluded epochs corresponding to trials with errors and or to trials with vocal onsets starting before 600ms or after 2300ms relative to the onset of the visual stimulus. At the end of the preprocessing, a mean of 90 epochs by participants (range = [16-219]) were excluded (i.e., 20% of trials).

We extracted both stimulus-aligned and response-aligned epochs so as to be able to capture effects that arise both early after picture onset and shortly before articulation. Stimulus-aligned ERPs capture processes that are aligned on the presentation of the stimulus, i.e., early encoding processes. Response-aligned ERPs on the other hand are particularly appropriate to target processes that are aligned on the vocal response, and that reflect later encoding processes (e.g., Bürki et al., 2015; Lancheros et al., 2020; Laganaro et al., 2013). We therefore created both a stimulus-locked and response-locked data set for each participant (e.g., Krott et al., 2019; Laganaro et al., 2012; Wong et al., 2017). The stimulus locked data set was created by segmenting epochs from -200ms to 500ms before and after visual stimulus onset (distractor-picture pair, distractors were presented with a stimulus-onset asynchrony of 0ms). Epochs of this data set were baseline corrected by using the mean of the signals between -200ms and 0ms relative to visual stimulus onset. The response locked data set was created by segmenting epochs from -600ms to -100ms relative to vocal response onsets. No baseline correction was applied to this data set. The raw and preprocessed EEG files are publicly available and can be found on OSF (https://osf.io/svjh5/).

**Analyses and results**

*Picture-word interference task: picture naming latencies*

Trials with incorrect responses on the picture-word interference task were removed from further analyses (n = 1150, 6%). Most errors occurred because the participants produced a word other than the intended word (n = 756, 62% of errors), were dysfluent (hesitations, false starts, n = 229, 19% of errors) or did not provide a response. Response latencies for each trial (with a correct response) were defined as the time between the onset of the picture presentation and the onset of the vocal response, and these were measured manually based on the spectrogram and oscillogram using the *Praat* software (Boersma & Weenink, 2014). The resulting dataset can be accessed here: https://osf.io/svjh5/. There were 58 trials for which the response of the participant or its onset could not clearly be defined (0.3% of the data); these trials were disregarded. We additionally removed 1791 trials that corresponded to epochs with artefacts in the EEG data, as well as 1044 trials with naming latencies below 600ms, such that the analysis of the response times and the analysis of the EEG signal are based on the same set of trials[3].

We analyzed response latencies with linear mixed-effects models (e.g., Goldstein, 1987) using the statistical software R (R core team, 2019) and the library lme4 (Bates et al., 2015). We estimated p-values with the lmerTest package (Kuznetsova et al., 2017), which uses the Satterthwaite method. Graphs were created using the package ggplot2 (Wickham, 2009). The scripts and datasets to reproduce these analyses are available publicly on OSF (https://osf.io/svjh5/). To facilitate replicability of our findings, we distinguish between planned and exploratory analyses. Planned analyses are those which were planned prior to data collection, whereas exploratory analyses were conducted after data collection or after initial analyses had been done.

---

[3] If trials with response times below 600ms were to be included, the ERP dataset would contain signal recorded during articulation. Given the difference in naming times across conditions, there would further be an imbalance in the number of epochs recorded during the articulation across conditions.

*Analysis 1. Group level analysis, replication of classical picture-word interference effects (planned)*

The first statistical model was conducted to determine whether the experiment could replicate the phonological facilitation, semantic interference, and general interference effects. Contrasts were determined manually (see Schad et al., 2020) such that the intercept represents the grand mean, and the four contrasts represent respectively the difference between phonologically related and unrelated trials (i.e., phonological contrast), the difference between semantically related and unrelated trials (semantic contrast), the difference between trials in the phonologically unrelated and baseline conditions (general interference contrast 1) and the difference between trials in the semantically unrelated and baseline conditions (general interference contrast 2). Related trials were coded as +1 and unrelated trials as -1. Trials in the baseline condition were coded as 1. The mean of all naming times was 888 ms ($SD = 143$, range = [600 – 2300 ms]), and average naming times for a given participant ranged between 742 ms and 1129 ms, and for items between 785 ms and 1056 ms. The mean naming latencies in each condition are presented in Figure .

Models had by-participant and by-item random intercepts and slopes for each contrast, but no correlations between intercepts and slopes. The model was run twice: once with all data points and once without outliers. Outliers were defined as any absolute values of the scaled residuals that were greater than 2.5. The outlier corrected model output is reported in Table 1. Note that the pattern of results is not different when these outliers are kept in the analysis. These results replicate the phonological facilitation, semantic interference, and general interference effects. Note that the same pattern of results is found when the naming latencies are log transformed.

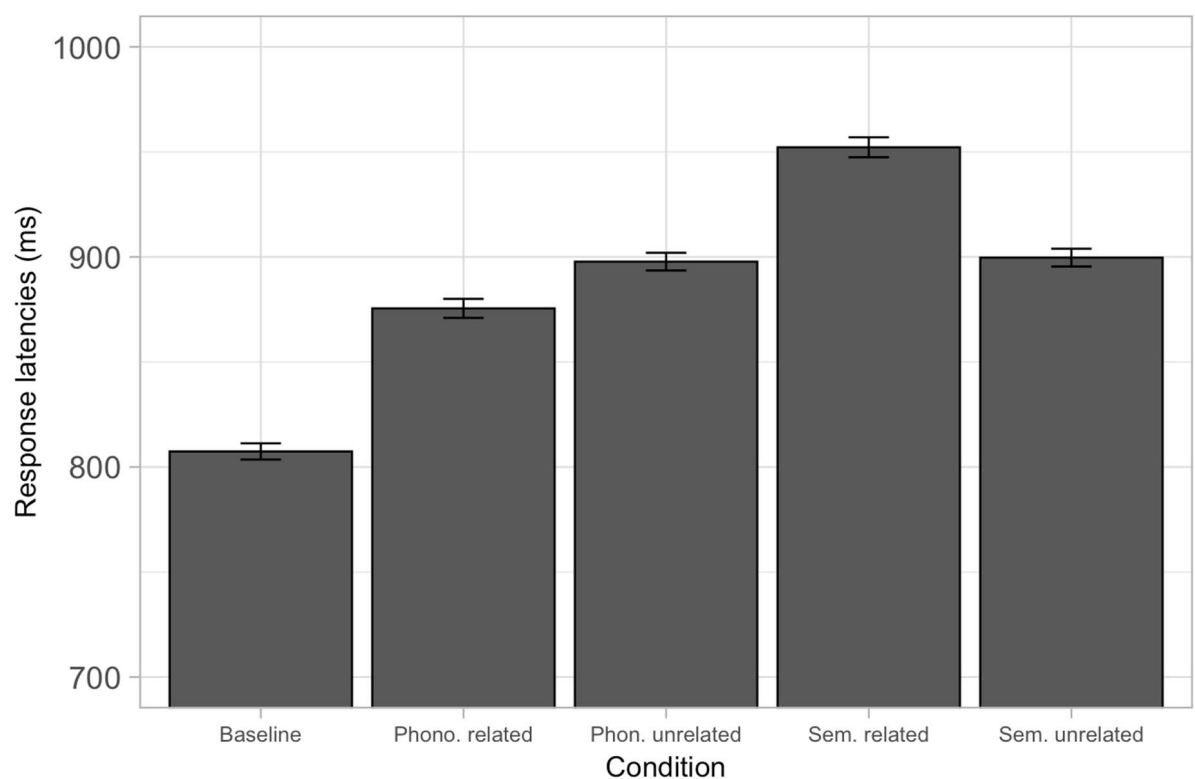

*Figure 2.* Picture-word-interference task: observed mean naming latencies and standard errors (values are adjusted for within-Participant designs following Morey, 2008) for each condition.

Table 1. Output of Analysis 1. Mixed-effects models testing the phonological, semantic, and general interference contrasts.

| | **Dependent Variable** | | | |
|---|---|---|---|---|
| *Predictors* | *Estimates* | *CI* | *t* | *p* |
| Intercept | 860.81 | 833.43 – 888.19 | 61.62 | <0.001 |
| Baseline - Unrelated (phon.) | 88.42 | 79.99 – 96.85 | 20.56 | <0.001 |
| Baseline - Unrelated (sem.) | 86.75 | 76.59 – 96.91 | 16.74 | <0.001 |
| Phonological contrast | -27.80 | -41.25 – -14.35 | -4.05 | <0.001 |
| Semantic contrast | 50.53 | 35.50 – 65.57 | 6.59 | <0.001 |
| N $_{Picture}$ | 90 | | | |

| | |
|---|---|
| N $_{Participant\_ID}$ | 45 |
| Marginal $R^2$ / Conditional $R^2$ | 0.052 / 0.326 |

*Analysis 2. Variability in immediate and delayed picture naming latencies (planned)*

The aim of this analysis was to determine how much of the variability in response times across participants can be accounted for by late articulatory execution processes. Recall that the delayed naming task is thought to represent late articulatory processes, while the immediate naming task (the picture-word interference task) is thought to reflect all stages of word production (Laganaro & Alario, 2006). The only trials that were eliminated from this analysis were trials with incorrect responses on either task and trials on the delayed task that had a negative reaction time (n = 323). Trials on the delayed task had a negative response time if the response was given prior to the response cue. To obtain a descriptive measure of between-participant variability in the immediate (in the baseline condition) and delayed naming tasks that would not be influenced by item variability or within-participant variability, we calculated the mean of the reaction times for each participant on each task (immediate: *M* = 791 ms, *SD* = 102 ms; delayed: *M* = 458 ms, *SD* = 105 ms). By-participant mean reaction times were submitted to a paired Levene's test (two-sided) using the levene.Var.test function from the PairedData package (Champely, 2018) in R to test for heterogeneity of variance in the two tasks. This function calculates a paired t-test on the absolute deviations from the mean or median (we used the median, immediate: *M* = 77 ms, *SD* = 69 ms; delayed: *M* = 82 ms, *SD* = 71 ms). We found no significant difference in between-participant variability between the two tasks, *t*(44) = -0.41, *p* = 0.69 (see Figure 3A-B).

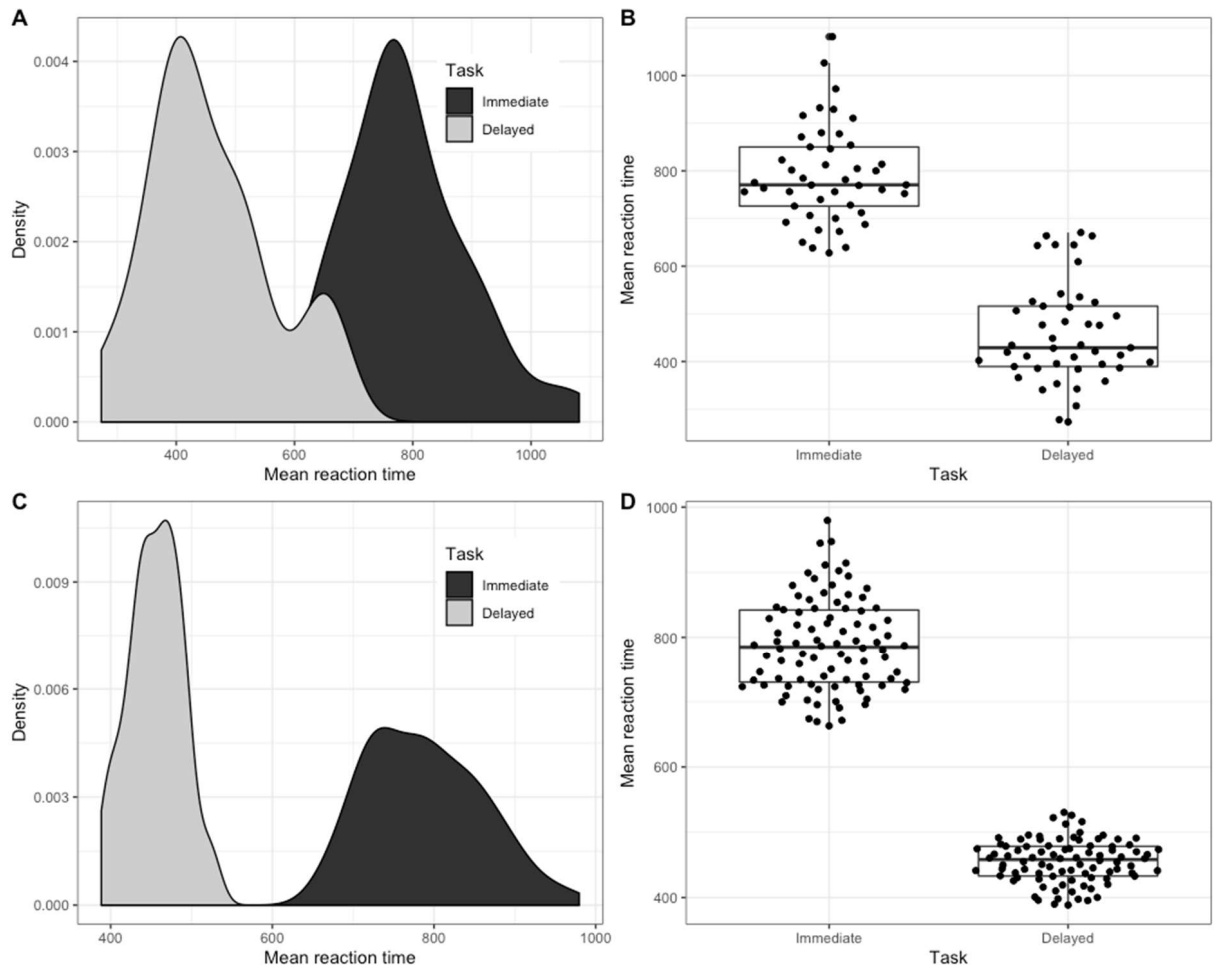

Figure 3. A. Density plot and B. boxplot showing the distributions of by-participant mean reaction times in each task. C. Density plot and D. boxplot showing the distributions of by-item mean reaction times in each task. Similar variance was observed between the two tasks for participant analysis, but variance differed in the two tasks (greater variance in immediate task than delayed) for item analyses.

If more between-participant variability emerges in late encoding processes, we would additionally expect the mean reaction times for each participant to be well correlated between the immediate (i.e., baseline condition with no distractors) and delayed naming tasks, and this is indeed what we found, $r = 0.58$ [95% CI 0.34-0.74], $p < 0.001$.

By comparison, certain properties of words (e.g., frequency) have been shown to influence word-naming latencies at earlier stages, such as lexical selection or phonological encoding (e.g., Alario et al., 2002; Levelt, 1999). If the immediate task reflects all stages of word production and the delayed task reflects only later stages, we would expect to see more

between-*item* variability in the immediate task compared to the delayed task (by-item descriptive statistics, immediate: *M* = 789 ms, *SD* = 71 ms; delayed: *M* = 455 ms, *SD* = 33 ms). We used the same procedure described above to test for differences in between-item variance between the two tasks and found that the variance in the immediate task (absolute deviations from the median, *M* = 58 ms, *SD* = 40 ms) was greater than the delayed task (absolute deviations from the median, *M* = 27 ms, *SD* = 19 ms), *t*(89) = 6.94, *p* < .001. We also would not expect the mean reaction times for each picture in the two tasks to be correlated; however, we did find a modest correlation between the two tasks, *r* = 0.23 [95% CI 0.2-0.42], *p* = 0.03. To test whether the correlation between the by-participant reaction times in the two tasks was stronger than the correlation between the by-item reaction times, we used the Fisher z-transformation to statistically compare the two correlations. This test revealed that the correlation between the by-participant reaction times for the two tasks was indeed stronger than the by-item correlation, *z* = 2.28 [95% CI .05-.61], *p* = .02.

These analyses do not provide support for the hypothesis that the variability across participants is greater in naming latencies that presumably reflect all encoding process, from picture recognition to initiation of articulation (immediate picture naming task), than in naming latencies that presumably only reflect execution processes (delayed picture naming task). These results are compatible with the hypothesis that a large part of differences across individuals in naming latencies actually arise in late encoding processes.

*ERPs in picture-word interference task*

*Analysis 1. Group-level effects of distractor condition (planned)*

The aim of this analysis was to assess the electrophysiological signature of the semantic interference, phonological facilitation, and general interference effect at the group level. For these and all other analyses described below, we relied on mass univariate analysis, testing for significant differences over the whole space of electrodes and every time point (e.g. Pernet et al., 2015), correcting for multiple comparison by applying the threshold free cluster enhancement (TFCE) procedure (Mensen et al., 2017; Mensen & Khatami, 2013; Smith & Nichols, 2009). Plots displaying group-level results of the semantic interference and phonological facilitation effects are displayed in Appendix 3, and results for the general interference effect are presented in supplementary materials found at https://osf.io/svjh5/. As described in detail in the EEG acquisition and preprocessing section, we conducted stimulus-locked analyses to test for early effects that were aligned on the stimulus onset and response-locked analyses to capture later effects aligned on the vocal response.

*Semantic interference effect.* In the stimulus-locked ERPs, the semantically related and semantically unrelated conditions differ between about 390 and 460 ms after picture onset on a set of posterior-central electrodes, with more positive values for the related condition than the unrelated, as can be seen on the plot with values on the electrode Pz (Appendix 3, Figure 1). In the response-locked ERPs, the two conditions differ from 420 ms to 100 ms before speech onset (i.e., last sample of analyzed time window). The effect starts on bilateral posterior electrodes and is subsequently distributed over the whole scalp, with a dissociation between fronto-central electrodes displaying less negative values for the related condition, and posterior electrodes displaying more negative values for the unrelated condition (Appendix 3, Figure 2).

*Phonological facilitation effect.* In the stimulus-locked ERPs, the phonologically related and unrelated conditions differ between about 420 and 500 ms on a set of posterior electrodes, mostly right lateralized, displaying less negative values in the related condition (Appendix 3, Figure 3). In the response-locked analysis, phonologically related and phonologically unrelated conditions differed statistically as soon as 600 ms before vocal onset on the posterior area, with less negative values in the unrelated condition (Appendix 3, Figure 4).

*Analysis 2. Single-subject analysis and correlations with cognitive measures and response times (planned)*

The aim of this analysis was to determine the time course of the semantic interference and phonological facilitation effects at the individual level and examine to what extent these can be related to participants' response times and performance on cognitive tasks. We examined the test contrasts for each participant (first level analysis in LIMO) and corrected for multiple comparisons using TFCE. For the semantic contrast, 0 out of 45 participants showed a significant effect in the stimulus-locked ERPs, and only six participants had an effect after correction in the response-aligned ERPs. For the phonological contrast, two and four participants showed a significant effect in the stimulus-aligned and response-aligned analyses, respectively. Given the small number of participants where effects could be detected, we were not in a position to correlate the time course of these effects with any other measure. Instead, we performed exploratory analyses, to be described below.

*Analysis 3. Experimental effects in fast versus slow speakers (exploratory)*

This analysis was inspired by Laganaro et al. (2012). These authors divided their participants into groups according to their response speed and performed an analysis for each speed group. With this analysis we can test the hypothesis that the time course of experimental effects depends on naming times. We first performed power analyses for each contrast to determine how many participants were needed to obtain at least 80% power to detect an effect of naming times. The power analyses were performed using the simR library (Green & MacLeod, 2016), a package that allows computing power functions for mixed-effects models using simulations.

These analyses confirm that with half the participants, we still have a high probability of detecting the semantic interference and general interference effects (100% power reached with less than 15 participants). For the phonological contrast, 23 participants (half of our sample) are needed to achieve 80% power. We are aware that these power analyses inform us on the likelihood of detecting an effect in the behavioral data (reaction times), not in the ERPs. However, the experimental effects we are studying with EEG are assumed to reflect behavioral effects, and we did not have enough information to compute the power for the EEG analyses. Therefore, for each contrast, we did a median split on participants' median reaction times of the conditions that made up that contrast to divide participants into two groups of fast and slow responders. For the semantic contrast, for example, we calculated each participant's median reaction time for the semantically related and unrelated trials and performed the median split on those values. In the present study, we use the semantic interference and phonological facilitation effects as markers of encoding processes in word production to test whether inter-individual differences in the speed of picture naming influence ERP effects, which is why we split participants into speed groups using only trials from the condition of interest (rather than, for example, using reaction times from the baseline condition). Investigating inter-individual differences in the *magnitude* of the semantic interference and phonological facilitation effects is outside the scope of the present study.

*Semantic interference effect.* In the stimulus-locked analysis of the semantic contrast, slow participants (n = 23, mean reaction times = 1003 ms, SD = 79 ms, range = [907 - 1161ms]) showed an effect on a small cluster after correction for multiple comparisons around 250-260 ms after picture presentation on right-lateralized posterior electrodes with more positive values for the semantically related condition (Figure 4). For fast participants (n = 22, mean reaction times = 844 ms, SD = 46, range = [778 – 920 ms]), a difference is found between 425 and 445 ms after picture onset on posterior electrodes with more positive values for the semantically related condition (Figure 5). If the semantic interference effect reflects lexical access and lexical access occurs later in slower speakers (e.g., Laganaro et al., 2012), we would have expected to see an effect in an earlier time window for fast than for slow participants. Assuming that we consider the very small effect in slow participant as meaningful, the data seem to show the reverse pattern.

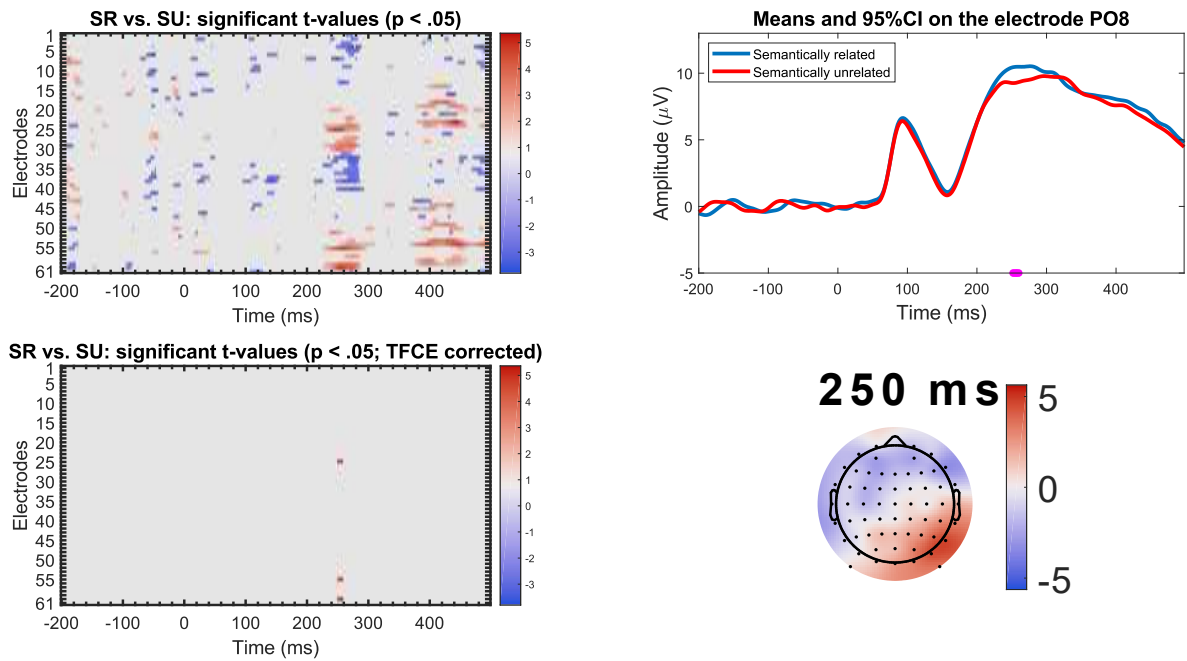

*Figure 4: Results of mass univariate analysis for the comparison between the semantically related (SR) and unrelated (SU) trials for the stimulus-locked ERPs of slow participants (n = 23), with significant t-values before correction (top left); significant t-values after correction with TFCE (bottom left); mean amplitude at electrode PO8 (electrode where t-value is maximal) for the semantically related and unrelated trials, significant time points (after TFCE correction) underlined in purple (top right); Topographic map of t-values before correction at 250 ms (bottom right).*

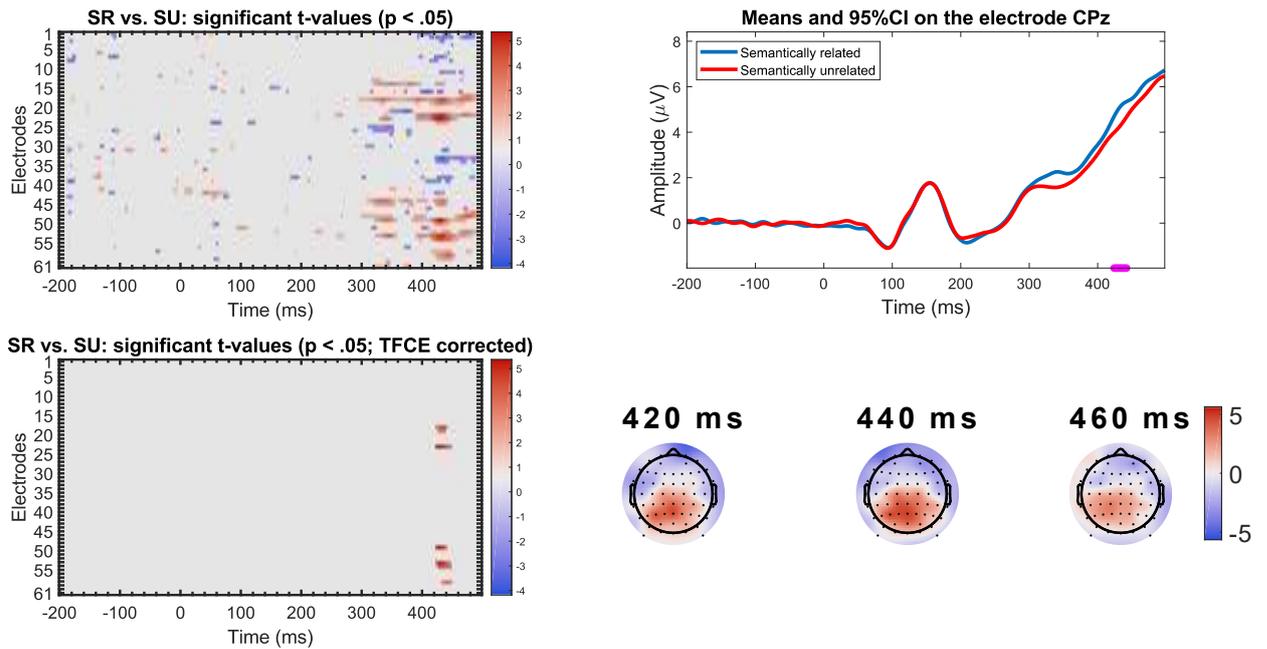

*Figure 5: Results of mass univariate analysis for the comparison between the semantically related (SR) and unrelated (SU) trials for the stimulus-locked ERPs of fast participants (n = 22), with significant t-values before correction (top left); significant t-values after correction with TFCE (bottom left); mean amplitude at electrode CPz (electrode where t-value is maximal) for the semantically related and unrelated trials, significant time points (after TFCE correction) underlined in purple (top right); Topographic map of t-values before correction (bottom right).*

In the response-locked analysis, we observed an effect in different time windows depending on the speed group. For the slow participants (n = 23), significant differences can be found as early as 500 ms before response onset on a set of posterior electrodes with less negative values for the semantically unrelated condition (Figure 6). Starting around 410 ms prior to the vocal response, a difference in conditions is found, with bilateral frontal electrodes showing less negative values for the semantically related condition and posterior electrodes showing less negative values for the semantically unrelated condition. The localization and direction of the effect is similar for fast participants (n = 22), except the effect starts closer to the response than for slow participants, around 220 ms before response onset (Figure 7). This difference between groups is as expected if slow and fast participants differ in the last encoding stages before the onset of articulatory movements.

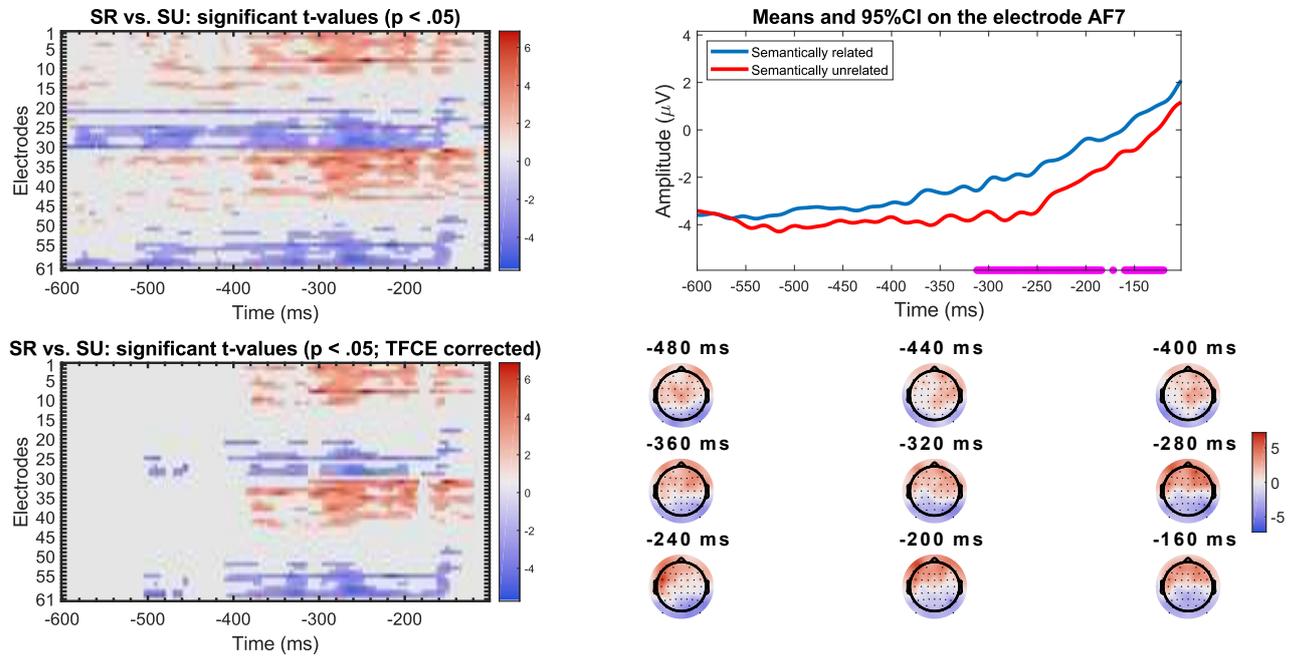

*Figure 6. Results of mass univariate analysis for the comparison between semantically related (SR) and unrelated (SU) trials for the response-locked ERPs in slow participants (n = 23) with significant t-values before correction (top left); significant t-values after correction with TFCE (bottom left); mean amplitude at electrode AF7 (electrode where t-value is maximal) for the semantically related and unrelated trials, significant time points (after TFCE correction) underlined in purple (top right); Topographic map of t-values before correction (bottom right).*

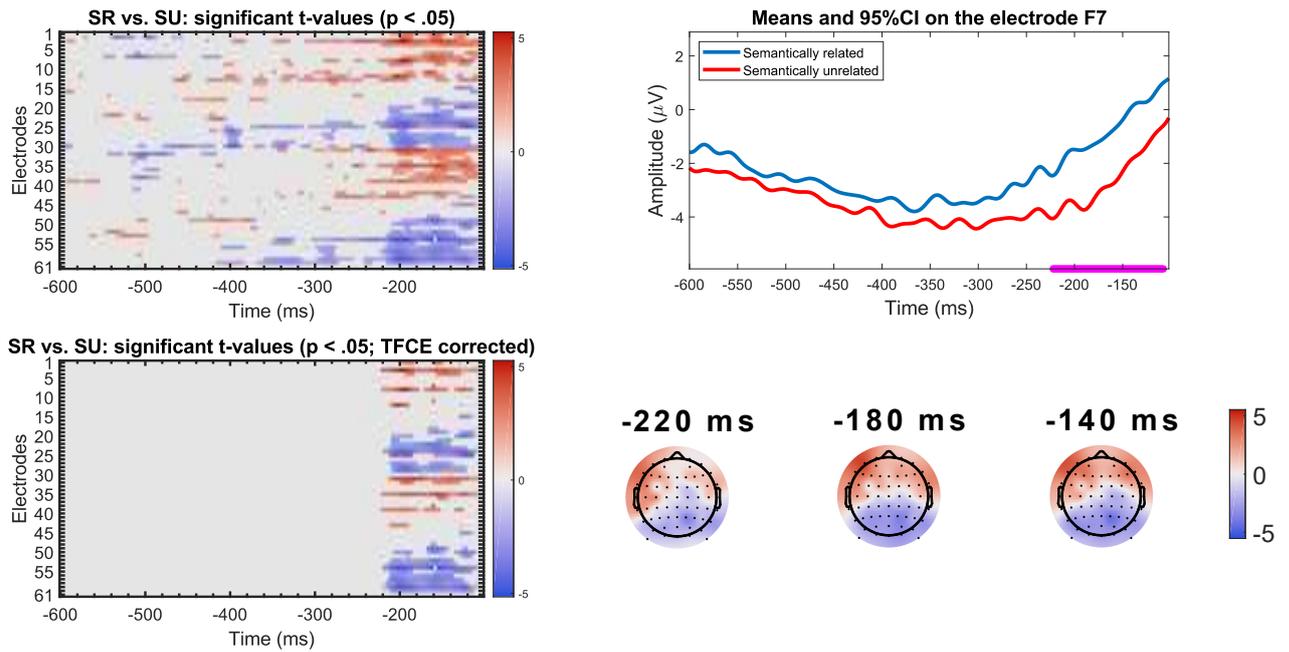

*Figure 7.* Results of mass univariate analysis for the comparison between semantically related (SR) and unrelated (SU) trials for the response-locked ERPs in fast participants (n = 22) with significant t-values before correction (top left); significant t-values after correction with TFCE (bottom left); mean amplitude at electrode F7 (electrode where t-value is maximal) for the semantically related and unrelated trials, significant time points (after TFCE correction) underlined in purple (top right); Topographic map of t-values before correction (bottom right).

*Phonological facilitation effect.* For the phonological contrast, no time point remained significant after correction for multiple comparisons for slow (n = 23, mean reaction times = 954 ms, SD = 79, range = [843 – 1131 ms]) or for fast speakers (n = 22, mean reaction times = 813 ms, SD = 45 ms, range = [720 – 883 ms]), in the response-locked or in the stimulus-locked analyses.

*Analysis 5. Time course of experimental effects in varying samples of participants (exploratory)*

The comparison of experimental effects across speed groups suggests that the timing of effects depends on participants' naming speed. When collecting data for an experiment, we use random samples of participants, with no a priori information on the speed with which our participants can name pictures. In this last analysis, we examined how the results of the group analyses vary with different samples of participants.

For each contrast (phonological and semantic), and following our power analyses on naming times, we took 20 random samples of 23 participants. For each sample, we performed a mass univariate analysis on response-locked and stimulus-locked ERPs, using TFCE to correct for multiple comparisons. We then visualized changes in the results across samples. The results of these four analyses are illustrated in Figure 8. Only 5% of samples showed effects at around 240 ms after stimulus onset in the stimulus-locked analysis of the semantic contrast (Figure 8A). The most robust results were found in the response-locked analyses for the semantic contrast, with some electrodes showing a significant effect at about 100-200 ms before word onset in 80-90% of the samples (Figure 8B). No samples showed significant effects in the stimulus-locked ERPs for the phonological contrast (Figure 8C), and only 20-25% of samples showed significant effects around 580 ms before word onset in the response-locked analysis for the phonological contrast (Figure 8D). These results suggest that the time course of effects—but also whether an effect is observed at all—depends on the sample of participants.

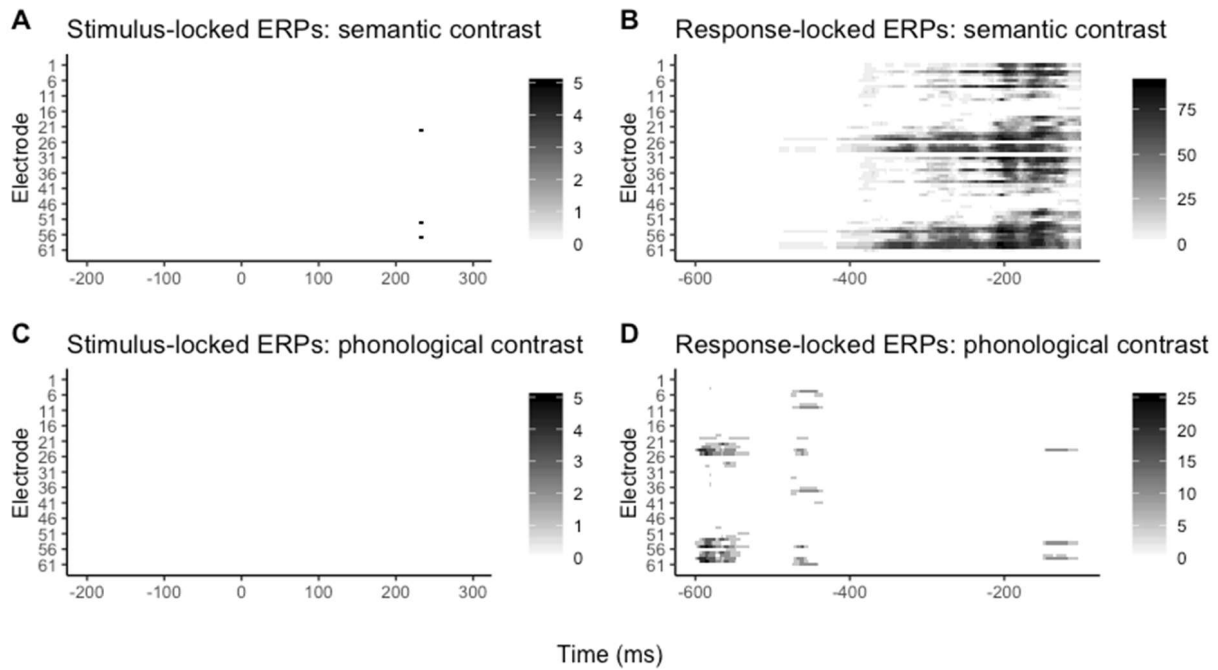

*Figure 8.* Results of mass univariate analyses for 20 random samples of 23 participants each. Note the difference in scales among plots. Plots display the percentage of time an electrode and time point are significant after TFCE correction on 23 participants ($p < .05$) for the A. stimulus-locked analysis for the semantic contrast, B. response-locked analysis for the semantic contrast, C. stimulus-locked analysis for the phonological contrast, and D. response-locked analysis for the phonological contrast.

**General discussion**

The present study investigated inter-individual variability in the time needed to prepare a vocal response for word production. Our first aim was to determine whether this variability could be associated with specific encoding processes, and our second aim was to examine the impact of inter-individual differences in the timing of encoding processes on group results in EEG studies.

*Locus of inter-individual differences in naming times*

As expected, participants' naming times in the present study were highly variable; mean picture-naming latencies ranged from 740 to 1130 ms. The slowest participant needed on average 390 additional milliseconds to perform the task. Where do these differences come from? Are slower participants slower in all encoding process, from conceptualization to execution, or do they need more time only in a subset of encoding processes? Several findings from the present study concur to support the hypothesis that a large part of inter-individual differences in response times originates in late encoding processes. The first source of evidence comes from the comparison of variability of individual naming times in the immediate naming task (baseline condition) and the delayed naming task. In a delayed naming task, participants have time to encode their response up to the phonetic level (at least for the first syllable of the response) before the initiation of the response. As a result, the naming times are assumed to mostly reflect the processes involved in the execution of articulatory gestures (Laganaro & Alario, 2006). Under this assumption, inter-individual variability in delayed response times reflects inter-individual variability in late execution processes. By contrast, naming times in an immediate naming task reflect the sum of all encoding processes, from conceptualization to articulation.

In the current study, we found that the between-participant variability was similar in the delayed and immediate naming tasks. Moreover, the correlation between the average by-participant response times of the two tasks was rather high, providing additional support for the hypothesis that a large part of inter-individual differences in response times arises in late encoding processes. We also found that in comparison, the correlation between the by-item mean response times in the two tasks was weaker. We know from previous studies that the properties of the words or of the pictures (frequency, age of acquisition, name agreement, see Perret & Bonin, 2019, for a recent meta-analysis of the variables that influence picture naming latencies) influence naming times. Unlike for between-*participant* variability, we have a fairly good knowledge of the origin of naming time differences across *items*. Many variables that are known to influence naming latencies have been associated with early encoding processes, namely conceptualization, mapping between picture and concept, lexical access, or phonological encoding. The weaker correlation between the mean naming times in the immediate and delayed tasks for the items is as expected if some of the variability across items arises in early encoding processes.

The EEG data also provide support for the hypothesis that at least part of the variability across participants arises in late encoding processes. For example, the difference between semantically related and unrelated trials is informative on the locus of inter-individual differences. In the whole-group analyses, the difference is mostly found in a late time window (between about 400 and 500 ms post-picture onset in the stimulus-aligned analyses) and is mostly aligned on the vocal responses. The timing of this effect and in particular the observation that the effect is more aligned on the response times than on the presentation of the picture favors a post-lexical locus of this effect (though our results do not necessarily exclude the possibility that lexical selection is not finalized until late encoding stages, see cascading models of word production, e.g., Dell, 2013; Peterson & Savoy, 1998; see also Lorenz et al., in press). This timing is compatible with accounts that situate this effect during the phonetic encoding process. According to the response-exclusion account, in a picture-word interference task, participants process the distractor and store its corresponding articulatory programs in a pre-response buffer. In order for the target word to be produced, the buffer must first be cleared. The suppression of the non-target response from the buffer depends on the relevance of this response. Distractors of the same semantic category are more relevant and therefore take longer to be suppressed (see for instance Mahon et al., 2007; but see for instance Abdel Rahman & Aristei, 2010). That we find more evidence for a late than an early locus of the semantic interference effect does not speak to whether part of the semantic interference effect also occurs during lexical access. Other EEG studies do report effects in time windows that are compatible with a lexical locus (e.g., Rose et al., 2019). The smaller difference we observe between related and unrelated trials in the stimulus-aligned ERPs could be taken to reflect a lexical locus, or the suppression of the distractor from the response buffer in the shorter trials/faster participants. Crucially, the comparison between slow and fast participants reveals that the effect in the response-aligned ERPs starts earlier in the slow as opposed to the fast participants (about 200 ms earlier) but lasts until the end of the analyzed time period for the two groups. Interestingly, the naming times are 135 ms shorter in the fast than in the slow participants. These results are as expected if a substantial part of the variability across participants originates in the last encoding processes before the onset of articulation. Again, this finding does not rule out the hypothesis that inter-individual variability also arises at earlier time points. The analysis performed on random samples suggests that the difference in amplitude between the related and unrelated trials in the stimulus-aligned ERPs is reduced when examined across samples. However, this analysis does not allow us to determine how this variability relates to participants' speed, and it is possible that the effect size in this time window is too small to be detected in smaller samples.

It is important to note that the picture-word interference paradigm does not reflect ecological word production. The mechanism assumed by the response-exclusion account of the semantic interference effect is specific to this paradigm and presumably irrelevant in more ecological tasks, where participants produce words in the absence of distractors. It remains to be assessed whether the differences across speed groups reflect processes that are specific to the picture-word interference paradigm (e.g., how quickly they can suppress the unwanted response) or processes that also occur in naming without distractors (e.g., how quickly the phonetic plans for the target word are loaded into the buffer).

In sum, whereas others before us have associated differences in response speed to earlier processes (e.g., lexical access, see Laganaro et al., 2012), our findings support the hypothesis that a large part of this variability arises in late encoding processes, including the phonetic encoding process and/or the execution stage. However, the absence of differences in earlier time windows does not allow us to support or reject the hypothesis that part of the variability in naming times also originates in earlier processes.

*Inter-individual differences and ERPs*

Our analyses suggest that inter-individual differences in naming times do influence the timing of experimental effects in the EEG signal. For example, the onset of the semantic interference effect differs by at least 200 ms across speed group. This result complements previous studies (e.g., Laganaro et al., 2012) in showing that response speed does not only generate differences in the brain signal across participants, it generates differences in the timing of experimental effects. This conclusion is further supported by the analysis with random samples of participants. This analysis suggests that the observed onset of an experimental effect varies with the sample of participants. Taken together, these analyses reveal that if technically, ERPs provide a high temporal resolution, variability across participants prevents their use to test precise hypotheses about the timing of encoding processes.

These findings could well explain the important differences reported across studies in language production research in the timing of specific effects. Take for instance the semantic interference effect. This effect has been examined in several EEG studies, and results are heterogeneous. A subset of studies report an absence of an effect (we restrict the discussion to amplitude analyses, Hirschfeld et al., 2008b; Piai et al., 2012), while others reported effects that vary in their timing and shape. For example, Blackford et al. (2012) found a smaller N400 that was centrally distributed for the semantically related condition as measured in a 350-550 ms time window. Cai et al. (2020) found evidence of the semantic interference effect around 320-360 ms after stimulus onset and even later (530-570 ms) when participants were under time pressure. Rose et al. (2019) found a stronger positive amplitude for related than unrelated trials that was visible first in posterior regions and extended to central regions between 234 and 480 ms. Roelofs, Piai, Garrido Rodriguez, and Chwilla (2016) found larger N400 amplitudes for semantically unrelated items compared to semantically related items.

Greenham et al. (2000) found a similar pattern, specifically larger amplitudes for semantically incongruent than congruent words, but with an N450 component. Using MEG (which allows for spatial precision as well as temporal precision), Dirani & Pylkkänen (2018) found the semantic interference effect at around 395-485 ms in frontal motor regions, and take this result to suggest that this effect occurs during motor planning. On the other hand, Dell'Acqua et al. (2010) found evidence of the semantic interference effect at a much earlier time window of around 100 ms after stimulus onset, mostly in left frontal and temporal electrodes. Abdel Rahman & Aristei (2010) also found interference effects at earlier time points, specifically as early as 200ms after stimulus onset in frontal and temporal electrodes. Rose and Abdel Rahman (2017) similarly found evidence of semantic interference at earlier time points, specifically between 100 and 150 ms, which they associate with a conceptual phase, and around 250 ms, which they associate with lexical access.

A few studies have reported both stimulus-aligned and response-aligned analyses, as was done in the current study. For example, a study by Krott et al. (2019) reports no difference in the response-locked analysis, but several clusters with differences in the stimulus-aligned ERPs, unlike in the present study. Wong et al. (2017) similarly found differences in semantically related and unrelated conditions in stimulus-locked ERPs. They found these differences in semantic relatedness in a time window between 275 and 400 ms after stimulus onset in anterior electrodes but did not find evidence of the semantic interference effect in response-aligned ERPs. Taken together, the findings discussed here suggest a wide range of variability in both timing and amplitude of ERP signatures of semantic interference effects.

Part of these differences in the results from the studies described above are likely due to differences in the analyses. For instance, many studies only examined the stimulus-aligned ERPs. In the present study, we analyzed both stimulus-aligned and response-aligned ERPs, and we performed the analysis with as few a priori restrictions as possible and used the most conservative correction for multiple comparisons. Unlike cluster-based permutation, TFCE provides information on significance for each time point and allows us to draw conclusions about the timing of effects (Sassenhagen & Draschkow, 2019). In contrast, other studies have averaged the signal in a priori (or a posteriori) defined time windows. Note also that in many of these studies, trials are included even though the response times were shorter than the analyzed time window. As a consequence, the EEG signal is recorded during articulation. Late effects in the stimulus-aligned time window might result from a difference in the number of trials capturing effects ongoing during articulation in the unrelated condition. The findings of the present study suggest that in addition to methodological differences, different results are expected as a result of inter-individual differences in processing speed.

In the present study, differences across participants are mostly found in late encoding processes. As a consequence, they warn against interpreting the timing of experimental effects in the response-aligned ERPs, but do not inform us on whether caution should also be taken in interpreting the timing of effects in the stimulus-aligned ERPs. We speculate however that the same warning applies there too. Even if participants' variability mostly arises in late encoding processes, variability is expected across items. We can therefore expect that the timing of experimental items in earlier time windows depends on the set of items used in the experiment.

If conclusions about the time course are necessary, the analysis that we performed with random samples can be performed to determine the confidence around the significant time point where the experimental effect is found. In the analysis of behavioral responses, accuracy or response times, the statistical model ensures that the effects generalize to other participants and items. This is for instance achieved by using crossed mixed-effects models with random terms for participants and items. In the analysis of ERPs, item is rarely considered a random variable (see also Bürki et al., 2018). More importantly, analyses of EEG, especially in the mass univariate framework, do not have a means to determine whether the timing of the effects generalize to other participants (or items). The sampling method coupled with a power analysis might prove useful in this respect but requires that the number of items and participants be large enough such that effects can be expected with smaller samples.

*Conclusion*

Inter-individual differences in behavioral tasks are apparent in everyday language tasks and at the experimental level, in the time needed to produce single words in response to a picture. In the present study, we tested whether individual variability emerges at earlier or late stages of word production. Data from a delayed naming task and EEG data suggest that a large part of the differences in response times we observe across participants arise in late encoding processes, presumably the phonetic encoding process. We further describe a novel way of analyzing ERP results, which informs us on the contribution of between-speaker variability to the group results.

Acknowledgements

This research was funded by the Deutsche Forschungsgemeinschaft (DFG, German Research Foundation) – project number 317633480 – SFB 1287, Project B05 (Bürki). The authors would like to thank Thea Villinger for her help in data collection and preprocessing.

Peterson, R. R., & Savoy, P. (1998). Lexical selection and phonological encoding during language production: Evidence for cascaded processing. *Journal of Experimental Psychology: Learning, Memory, and Cognition*, *24*(3), 539–557. https://doi.org/10.1037/0278-7393.24.3.539

Piai, V., Roelofs, A., Acheson, D. J., & Takashima, A. (2013). Attention for speaking: Neural substrates of general and specific mechanisms for monitoring and control. *Frontiers in Human Neuroscience*, *7*.

Piai, V., Roelofs, A., Jensen, O., Schoffelen, J.-M., & Bonnefond, M. (2014). Distinct Patterns of Brain Activity Characterise Lexical Activation and Competition in Spoken Word Production. *PLoS ONE*, *9*(2), e88674. https://doi.org/10.1371/journal.pone.0088674

Piai, V., Roelofs, A., & van der Meij, R. (2012). Event-related potentials and oscillatory brain responses associated with semantic and Stroop-like interference effects in overt naming. *Brain Research*, *1450*, 87–101. https://doi.org/10.1016/j.brainres.2012.02.050

Posnansky, C. J., & Rayner, K. (1977). Visual-feature and response components in a picture-word interference task with beginning and skilled readers. *Journal of Experimental Child Psychology*, *24*(3), 440–460. https://doi.org/10.1016/0022-0965(77)90090-X

R core team. (2014). *R: A Language and Environment for Statistical Computing*. R Foundation for Statistical Computing. http://www.R-project.org

Rabovsky, M., Schad, D., & Rahman, R. A. (2020). Semantic richness and density effects on language production: Electrophysiological and behavioral evidence. *bioRxiv*, 509000.

Rayner, K., & Posnansky, C. (1978). Stages of processing in word identification. *Journal of Experimental Psychology: General*, *107*(1), 64–80. https://doi.org/10.1037/0096-3445.107.1.64

Riès, S. K. (2016). Serial versus parallel neurobiological processes in language production: Comment on Munding, Dubarry, and Alario, 2015. *Language, Cognition and Neuroscience*, *31*(4), 476–479. https://doi.org/10.1080/23273798.2015.1117644

*Appendix 1. Material used in experiment*

|  | **Distractor type** | | | |
|---|---|---|---|---|
| **Target word** | **Semantically related** | **Semantically unrelated** | **Phonologically related** | **Phonologically unrelated** |
| Dusche (shower) | Badewanne (bathtub) | Frisbee (frisbee) | Duell (duel) | Banjo (banjo) |
| Knochen (bone) | Muskel (muscle) | Radar (radar) | Knospe (bud) | Salbe (ointment) |
| Leopard (leopard) | Tiger (tiger) | Kuchen (cake) | Lehm (clay) | Dimension (dimension) |
| Zwiebel (onion) | Knoblauch (garlic) | Stock (stick) | Zwiesel (zwiesel) | Antike (antiquity) |
| Mikroskop (microscope) | Fernglas (binoculars) | Strumpf (stocking) | Mikado (mikado) | Geschichte (story) |
| Spinne (spider) | Ameise (ant) | Duschgel (shower gel) | Spitzel (spy) | Duell (duel) |
| Kleeblatt (shamrock) | Schilf (reed) | Zange (pliers) | Klerus (clergy) | Hamster (hamster) |
| Bürste (brush) | Kamm (comb) | Rubin (ruby) | Bürde (burden) | Chaos (chaos) |
| Ellenbogen (elbow) | Knie (knee) | Postkarte (postcard) | Elfe (elf) | Feudalismus (feudalism) |
| Brief (letter) | Postkarte (postcard) | Wal (whale) | Brigade (brigade) | Taste (button) |
| Zeitung (newspaper) | Buch (book) | Herd (stove) | Zeichen (sign) | Trost (consolation) |
| Kerze (candle) | Fackel (torch) | Muskel (muscle) | Kerbe (notch) | Hafen (port) |
| Banane (banana) | Aprikose (apricot) | Streichholz (match) | Banjo (banjo) | Schwager (brother-in-law) |
| Diamant (diamond) | Rubin (ruby) | Mücke (mosquito) | Dialekt (dialect) | Kruste (crust) |
| Krücke (crutch) | Stock (stick) | Ameise (ant) | Kruste (crust) | Panther (panther) |
| Mond (moon) | Sonne (sun) | Huhn (chicken) | Monitor (monitor) | Biene (bee) |
| Koffer (suitcase) | Tasche (bag) | Stift (pen) | Koffein (caffeine) | Adel (nobility) |
| Hirsch (deer) | Elch (moose) | Säge (saw) | Hirn (brain) | Akt (act) |
| Kiwi (kiwi) | Zitrone (lemon) | Tasche (bag) | Kino (movie theater) | Armee (army) |
| Ratte (rat) | Maus (mouse) | Füller (pen) | Raster (grid) | Pickel (pimple) |
| Kühlschrank (refrigerator) | Herd (stove) | Saxophon (saxophone) | Kübel (bucket) | Zwiesel (zwiesel) |
| Geist (ghost) | Zombie (zombie) | Floß (raft) | Geier (vulture) | Konsum (consumption) |
| Olive (olive) | Bohne (bean) | Lineal (ruler) | Olympia (olympics) | Zirkus (circus) |
| Domino (domino) | Würfel (dice) | Papaya (papaya) | Dominanz (dominance) | Zeichen (sign) |
| Kokosnuss (coconut) | Papaya (papaya) | Deodorant (deodorant) | Kokain (cocaine) | Analyse (analysis) |
| Regen (rain) | Schnee (snow) | Mütze (cap) | Regel (rule) | Mikado (mikado) |
| Zirkel (divider) | Lineal (ruler) | Flasche (bottle) | Zirkus (circus) | Seide (silk) |
| Rock (skirt) | Hemd (shirt) | Gras (grass) | Rost (rust) | Hirn (brain) |

# INDIVIDUAL VARIABILITY IN LATE STAGES OF WORD PRODUCTION

| | | | | |
|---|---|---|---|---|
| Trompete (trumpet) | Saxophon (saxophone) | Schnee (snow) | Troll (troll) | Kokain (cocaine) |
| Kappe (cap) | Mütze (cap) | Würfel (dice) | Kapitän (captain) | Troll (troll) |
| Drucker (printer) | Scanner (scanner) | Zombie (zombie) | Druide (druid) | Hotel (hotel) |
| Bumerang (boomerang) | Frisbee (frisbee) | Aprikose (apricot) | Bulle (bull) | Druide (druid) |
| Hammer (hammer) | Zange (Pliers) | Traube (grape) | Hamster (hamster) | Bulle (bull) |
| Schnuller (pacifier) | Flasche (bottle) | Sessel (armchair) | Schnur (line) | Marine (navy) |
| Pinsel (brush) | Füller (pen) | Jacuzzi (jacuzzi) | Pickel (pimple) | Bote (messenger) |
| Bier (beer) | Wein (wine) | Kamm (comb) | Biene (bee) | Tante (aunt) |
| Salzstreuer (salt shaker) | Pfeffermühle (pepper grinder) | Pistole (pistol) | Salbe (ointment) | Pudel (poodle) |
| Schaufel (shovel) | Hacke (pickaxe) | Ziege (goat) | Schau (show) | Kerbe (notch) |
| Baguette (baguette) | Schrippe (bread roll) | Fuß (foot) | Bagger (excavator) | Hain (grove) |
| Hase (rabbit) | Wiesel (weasel) | Schilf (reed) | Hafen (port) | Rost (rust) |
| Bett (bed) | Sofa (sofa) | Granate (grenade) | Berg (mountain) | Tuba (tuba) |
| Chamäleon (chameleon) | Eidechse (lizard) | Pfeffermühle (pepper grinder) | Chaos (chaos) | Olympia (olympics) |
| Karotte (carrot) | Gurke (cucumber) | Eidechse (lizard) | Kardinal (cardinal) | Monitor (monitor) |
| Tunnel (tunnel) | Höhle (cave) | Harfe (harp) | Tuba (tuba) | Kanister (canister) |
| Kompass (compass) | Uhr (clock) | Melone (melon) | Konsum (consumption) | Beere (berry) |
| Dinosaurier (dinosaurs) | Mammut (mammoth) | Badewanne (bathtub) | Dimension (dimension) | Mediation (mediation) |
| Strauß (ostrich) | Huhn (chicken) | Knie (knee) | Strauch (shrub) | Kübel (bucket) |
| Faden (thread) | Wolle (wool) | Gurke (cucumber) | Fahne (banner) | Kirche (church) |
| Blume (flower) | Gras (grass) | Trophäe (trophy) | Bluse (blouse) | Pixel (pixel) |
| Parfüm (perfume) | Deodorant (deodorant) | Höhle (cave) | Park (park) | Brigade (brigade) |
| Kuh (cow) | Ziege (goat) | Dose (can) | Kuli (pen) | Fahne (banner) |
| Geschenk (gift) | Paket (package) | Sofa (sofa) | Geschichte (story) | Spitzel (spy) |
| Schmetterling (butterfly) | Libelle (dragonfly) | Marmelade (jam) | Schmerz (pain) | Dominanz (dominance) |
| Gitarre (guitar) | Harfe (harp) | Libelle (dragonfly) | Giraffe (giraffe) | Kreis (circle) |
| Tanne (fir) | Fichte (spruce) | Messer (knife) | Tante (aunt) | Regel (rule) |
| Pool (pool) | Jacuzzi (jacuzzi) | Gans (goose) | Pudel (poodle) | Ehre (honor) |
| Bombe (bomb) | Granate (grenade) | Wiesel (weasel) | Borste (bristle) | Filz (felt) |
| Antenne (antenna) | Radar (radar) | Jalousie (louvre) | Antike (antiquity) | Kardinal (cardinal) |
| Finger (finger) | Zeh (toe) | Puppe (doll) | Filz (felt) | Elfe (elf) |
| Vorhang (curtain) | Jalousie (louvre) | Mammut (mammoth) | Vorort (suburb) | Kredit (credit) |
| Thron (throne) | Sessel (armchair) | Hemd (shirt) | Trost (consolation) | Raster (grid) |
| Axt (axe) | Säge (saw) | Jeep (jeep) | Akt (act) | Lehm (clay) |
| Krebs (crab) | Hummer (lobster) | Uhr (clock) | Kredit (credit) | Schau (show) |
| Kanone (cannon) | Pistole (pistol) | Tablette (tablet) | Kanister (canister) | Vorort (suburb) |

# INDIVIDUAL VARIABILITY IN LATE STAGES OF WORD PRODUCTION

| | | | | |
|---|---|---|---|---|
| Ananas (pineapple) | Melone (melon) | Fernglas (binoculars) | Analyse (analysis) | Bluse (blouse) |
| Fliege (fly) | Mücke (mosquito) | Scanner (scanner) | Fliese (tile) | Dialekt (dialect) |
| Arm (arm) | Fuß (foot) | Wein (wine) | Armee (army) | Fliese (tile) |
| Rose (rose) | Tulpe (tulip) | Falke (falcon) | Rosine (raisin) | Fund (discovery) |
| Boot (boat) | Floß (raft) | Maus (mouse) | Bote (messenger) | Schnur (line) |
| Pizza (pizza) | Kuchen (cake) | Hacke (pickaxe) | Pixel (pixel) | Hantel (dumbbell) |
| Tasse (cup) | Dose (can) | Elch (moose) | Taste (button) | Schmerz (pain) |
| Fuchs (fox) | Wolf (wolf) | Zeh (toe) | Fund (discovery) | Gabe (gift) |
| Schwan (swan) | Gans (goose) | Zitrone (lemon) | Schwager (brother-in-law) | Borste (bristle) |
| Hai (shark) | Wal (whale) | Buch (book) | Hain (grove) | Berg (mountain) |
| Marionette (puppet) | Puppe (doll) | Knoblauch (garlic) | Marine (navy) | Koffein (caffeine) |
| Pille (pill) | Tablette (tablet) | Wolf (wolf) | Pilot (pilot) | Strauch (shrub) |
| Feuerzeug (lighter) | Streichholz (match) | Hummer (lobster) | Feudalismus (feudalism) | Kapitän (captain) |
| Handschuh (glove) | Strumpf (stocking) | Fichte (spruce) | Hantel (dumbbell) | Rosine (raisin) |
| Besen (broom) | Harke (rake) | Mandel (almond) | Beere (berry) | Geier (vulture) |
| Honig (honey) | Marmelade (jam) | Speer (spear) | Hotel (hotel) | Park (park) |
| Seife (soap) | Duschgel (shower gel) | Fackel (torch) | Seide (silk) | Klerus (clergy) |
| Panzer (tank) | Jeep (jeep) | Sonne (sun) | Panther (panther) | Boden (ground) |
| Kreide (chalk) | Stift (pen) | Tulpe (tulip) | Kreis (circle) | Mantra (mantra) |
| Mantel (coat) | Anorak (jacket) | Bohne (bean) | Mantra (mantra) | Bürde (burden) |
| Medaille (medal) | Trophäe (trophy) | Anorak (jacket) | Mediation (mediation) | Kuli (pen) |
| Gabel (fork) | Messer (knife) | Tiger (tiger) | Gabe (gift) | Knospe (bud) |
| Bogen (bow) | Speer (spear) | Wolle (wool) | Boden (ground) | Giraffe (giraffe) |
| Erdnuß (peanut) | Mandel (almond) | Harke (rake) | Ehre (honor) | Pilot (pilot) |
| Kirsche (cherry) | Traube (grape) | Paket (package) | Kirche (church) | Bagger (excavator) |
| Adler (eagle) | Falke (falcon) | Schrippe (bread roll) | Adel (nobility) | Kino (movie theater) |
| Aquarium (aquarium) | Käfig (cage) | Reis (rice) | Aquarell (watercolor) | Liste (list) |
| Drachen (kite) | Ballon (balloon) | Kabel (electric wire) | Draht (wire) | Kehle (throat) |
| Kartoffel (potato) | Reis (rice) | Augenbinde (blindfold) | Karton (carton) | Schlaufe (loop) |
| Kegel (pin) | Dart (darts) | Auge (eye) | Kehle (throat) | Mast (mast) |
| Lippen (lips) | Auge (eye) | Käfig (cage) | Liste (list) | Draht (wire) |
| Maske (mask) | Augenbinde (blindfold) | Kralle (claw) | Mast (mast) | Schnaps (schnapps) |
| Schlauch (hose) | Kabel (electric wire) | Dart (darts) | Schlaufe (loop) | Karton (carton) |
| Schnabel (beak) | Kralle (claw) | Ballon (balloon) | Schnaps (schnapps) | Aquarell (watercolor) |

*Appendix 2. Correspondence between electrode numbers on graphs and electrode names*

| Electrode | Number | Electrode | Number |
|---|---|---|---|
| Fp1 | 1 | AF3 | 32 |
| Fp2 | 2 | AF4 | 33 |
| F7 | 3 | AF8 | 34 |
| F3 | 4 | F5 | 35 |
| Fz | 5 | F1 | 36 |
| F4 | 6 | F2 | 37 |
| F8 | 7 | F6 | 38 |
| FC5 | 8 | FT7 | 39 |
| FC1 | 9 | FC3 | 40 |
| FC2 | 10 | FC4 | 41 |
| FC6 | 11 | FT8 | 42 |
| T7 | 12 | C5 | 43 |
| C3 | 13 | C1 | 44 |
| Cz | 14 | C2 | 45 |
| C4 | 15 | C6 | 46 |
| T8 | 16 | TP7 | 47 |
| CP5 | 17 | CP3 | 48 |
| CP1 | 18 | CPz | 49 |
| CP2 | 19 | CP4 | 50 |
| CP6 | 20 | TP8 | 51 |
| P7 | 21 | P5 | 52 |
| P3 | 22 | P1 | 53 |
| Pz | 23 | P2 | 54 |
| P4 | 24 | P6 | 55 |
| P8 | 25 | PO7 | 56 |
| PO9 | 26 | PO3 | 57 |
| O1 | 27 | POz | 58 |
| Oz | 28 | PO4 | 59 |
| O2 | 29 | PO8 | 60 |
| PO10 | 30 | FCz | 61 |
| AF7 | 31 | | |

*Appendix 3. ERP group results of semantic and phonological contrasts*

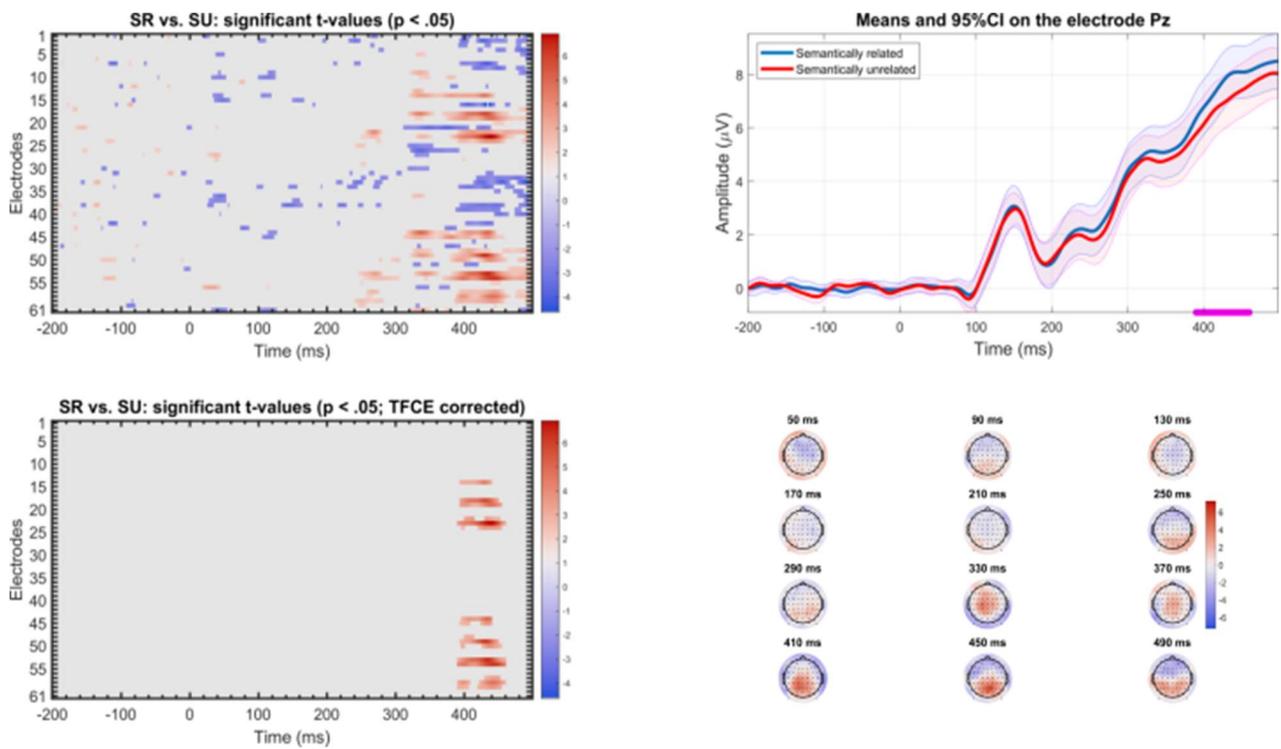

*Appendix 3, Figure 1. Results of the mass univariate analysis for the comparison between semantically related (SR) and semantically unrelated (SU) trials for the stimulus-locked ERPs, with significant t-values before correction (top left); significant t-values after correction with TFCE (bottom left); mean amplitude at electrode Pz (electrode where the t-value is maximal) for semantically related and unrelated conditions, significant time points (after TFCE correction) underlined in purple (top right); Topographic map of t-values before correction (bottom right).*

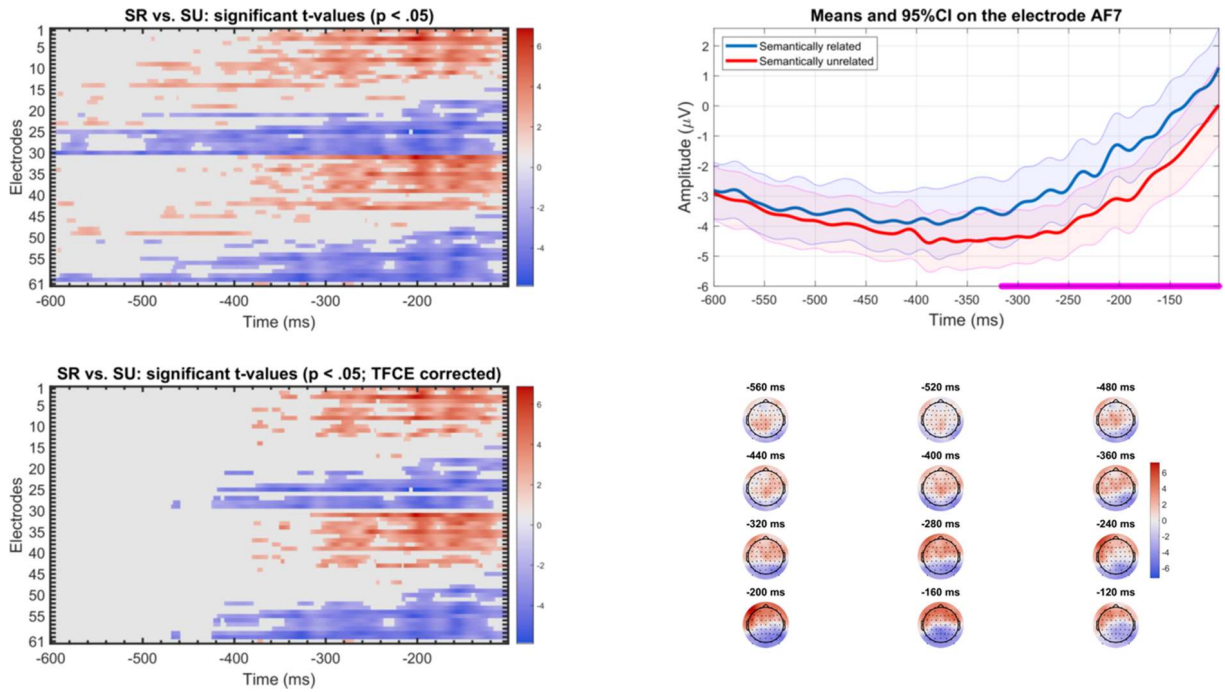

*Appendix 3, Figure 2.* Results of mass univariate analysis for the comparison between semantically related (SR) and semantically unrelated (SU) trials for the response-locked ERPs, with significant t-values before correction (top left); significant t-values after correction with TFCE (bottom left); mean amplitude at electrode AF7 (electrode where t-value is maximal) for semantically related and unrelated conditions, significant time points (after TFCE correction) underlined in purple (top right); Topographic map of t-values before correction (bottom right).

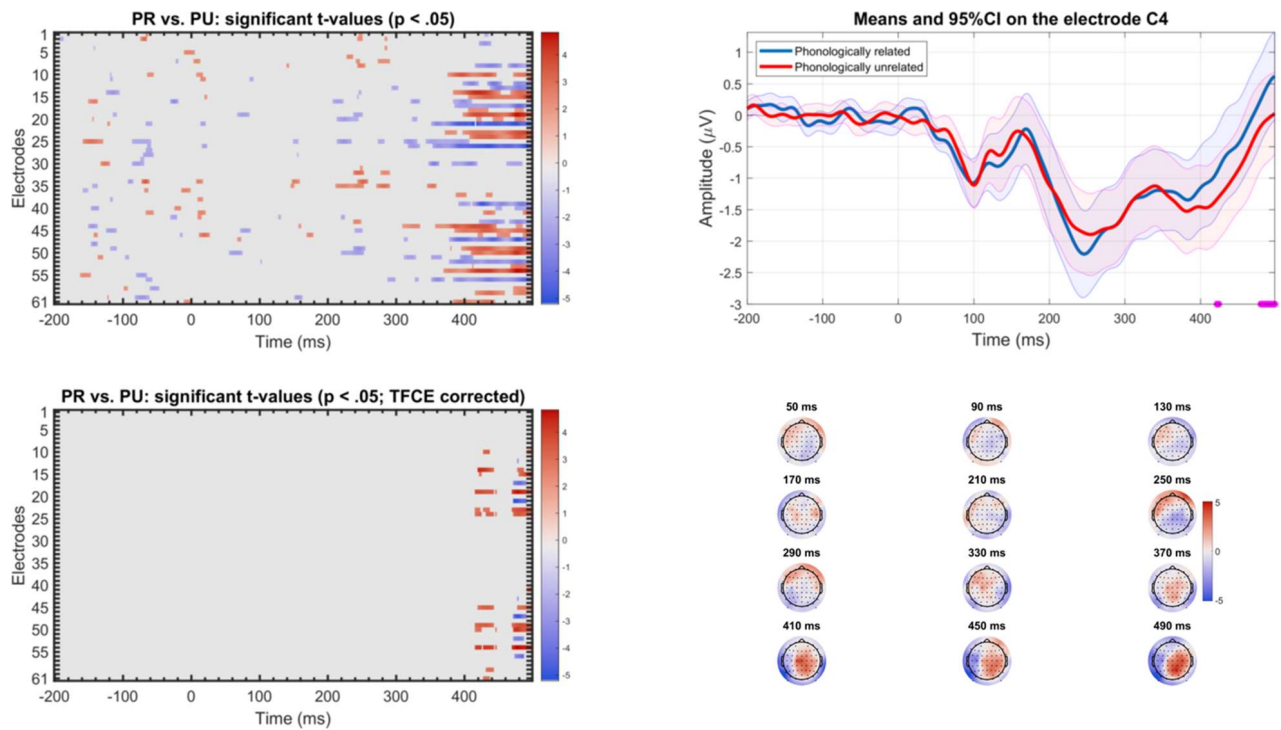

*Appendix 3, Figure 3.* Results of mass univariate analysis for the comparison between phonologically related (PR) and phonologically unrelated (PU) trials for the stimulus-locked ERPs, with significant t-values before correction (top left); significant t-values after correction with TFCE (bottom left); mean amplitude at electrode C4 (electrode where t-value is maximal) for phonologically related and unrelated conditions, significant time points (after TFCE correction) underlined in purple (top right); Topographic map of t-values before correction (bottom right).

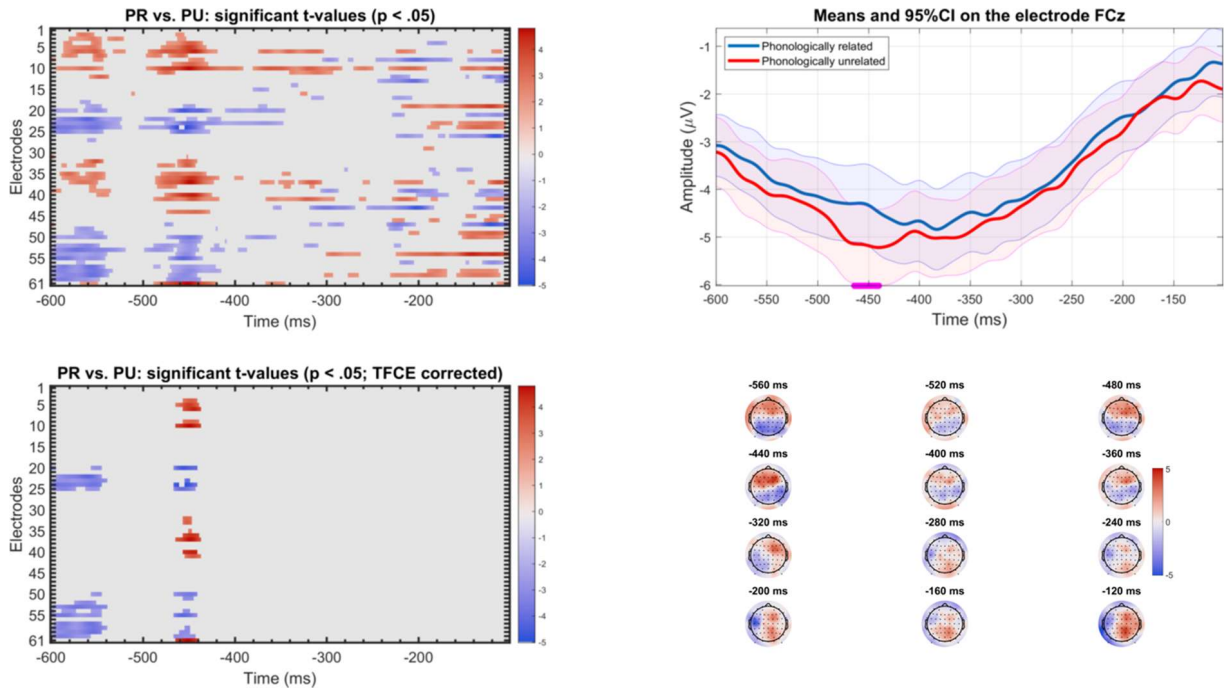

*Appendix 3, Figure 4.* Results of mass univariate analysis for the comparison between phonologically related (PR) and phonologically unrelated (PU) trials for the response-locked ERPs, with significant t-values before correction (top left); significant t-values after correction with TFCE (bottom left); mean amplitude at electrode FCz (electrode where t-value is maximal) for semantically related and unrelated conditions, significant time points (after TFCE correction) underlined in purple (top right); Topographic map of t-values before correction (bottom right).